\documentclass[a4paper,12pt]{article}

\usepackage[a4paper]{geometry}
\usepackage{caption}
\usepackage{epsfig}
\usepackage{amsmath}
\usepackage{xcolor}
\usepackage{hyperref}

\usepackage{amsthm}
\newtheorem{thm}{Theorem}
\newtheorem*{cor}{Corollary}
\usepackage{wasysym}
\usepackage[square,sort,comma,numbers]{natbib}

\title{The Scale Invariant Vacuum Paradigm:\\
\Large Main Results and Current Progress Review (Part II)}

\author{
Vesselin G.~Gueorguiev\footnote{e-mail: Vesselin at MailAPS.org}\\
{\it \small Institute for Advanced Physical Studies, Sofia, Bulgaria,}\\
{\it \small Ronin Institute for Independent Scholarship, NJ, USA}\\
\\
Andr\'e Maeder\footnote{e-mail: Andre.Meader at UniGe.ch}\\
{\it \small Geneva Observatory, University of Geneva, Switzerland}\\
}

\begin{document}
\date{\small \today}
\maketitle

\abstract{
We present a summary of the main results within the Scale Invariant Vacuum (SIV) paradigm 
based on the Weyl Integrable Geometry 
{(WIG) as an extension to the standard Einstein General Relativity (EGR)}.
After a brief review of the mathematical framework, 
where we also highlight the connection between the weak-field SIV equations 
and the notion of un-proper time parametrization within 
the reparametrization paradigm \cite{sym13030379},
we continue with the main results related to early Universe; that is,
applications to inflation \cite{SIV-Inflation'21}, Big Bang Nucleosynthesis \cite{VG&AM'23},
and the growth of the density fluctuations \cite{MaedGueor19}  within the SIV.
In the late time Universe the applications of the  SIV paradigm are related to
scale-invariant dynamics of galaxies, MOND, 
dark matter, and the dwarf spheroidals \cite{MaedGueor20b}
where one can find MOND to be a peculiar case of the SIV theory \cite{Maeder23}.
Finally, within the recent time epoch, we highlight that some of the  change in the length-of-the-day (LOD), 
about 0.92 cm/yr,  can be accounted for by SIV effects in the Earth-Moon system \cite{2022arXiv220413560M}.
}


\tableofcontents 

\section{Motivation}

The paper is a summary of the main results, as of midyear 2023, within the
Scale Invariant Vacuum (SIV) paradigm as related to the Weyl Integrable Geometry 
(WIG) as an extension to the standard Einstein General Relativity (EGR).
Our main goal is to present a condensed overview of the key results of the theory so far,
along with the latest  progress in applying the SIV paradigm to variety of physics phenomenon,
and in doing so to help the intellectually curious reader  gain some understanding as to where the 
paradigm has been tested and what is the success level of the inquiry. As such, the paper
follows closely our previous 2022 paper \cite{2022Univ....8..213G} that was based on 
the talk presented at the conference Alternative Gravities and Fundamental Cosmology, 
at the University of Szczecin, Poland in September 2021. Our initial presentation and 
its conference contribution were covering, back then, only four main results:
comparing the scale factor $a(t)$ within $\Lambda$CDM and SIV \cite{Maeder17a},
the growth of the density fluctuations within the SIV \cite{MaedGueor19},
the application to scale-invariant dynamics of galaxies \cite{MaedGueor20b},
and inflation of the early-universe within the SIV theory \cite{SIV-Inflation'21}.
Back then, our article layout was aiming for focusing on each of these four main results 
via highlighting its most relevant figure or equation. As a result each topic was covered via
one to two pages text preceded by short and concise description of the mathematical framework.

Here, we add a few new sections, one  on the possible differentiators 
of SIV from $\Lambda$CDM based on our earlier paper \cite{MaedGueor20a}, 
with a specific emphasis on the distance moduli as function of the redshift, along with 
three new topic sections related to the recent developments in the 
application of SIV paradigm since our previous summary paper in 2022 \cite{2022Univ....8..213G}.
The sections are on MOND as a peculiar case of the SIV theory \cite{MaedGueor20b},
local dynamical effects within SIV as pertained to the lunar recession \cite{2022arXiv220413560M},
and our latest study of the Big-Bang Nucleosynthesis (BBNS) within the SIV Paradigm  \cite{VG&AM'23}.

The paper begins with a general introduction to the problem of scale invariance and physical reality, 
as well as the similarities and differences between Einstein's general relativity and Weyl integrable geometry. 
Then we briefly review the mathematical framework of Weyl integrable geometry, 
Dirac co-calculus, and reparametrization invariance. 
We use the idea of reparametrization invariance \cite{sym13030379} 
to illustrate the corresponding equations of motion, rather than re-deriving the weak-field SIV results. 
The relevant discussion on reparametrization invariance is in 
Section \ref{sec:beyondEGR}  on the Consequences of Going beyond Einstein's General Relativity. 
This section precedes the brief review of the necessary results about 
the Scale Invariant Cosmology idea needed in the Section on Comparisons and Applications, 
where we highlight the main results related to the early and late Universe 
in the order seen in the table of contents and also discussed at the beginning of this section. 
The paper concludes with a section containing the Conclusions and Outlook for future research directions.

\subsection{Scale Invariance and Physical Reality}

The existence of a scale is related to the presence of physical connection and causality. 
The corresponding relationships are formulated as physical laws expressed in mathematical equations. 
Numerical factors in the formulas of physics laws change upon change of scale but maintain their mathematical form, 
thus exhibiting form-invariance. As a result, using consistent units is essential in physics and leads to 
powerful dimensional estimates of the order of magnitude of physical quantities based on a simple dimensional analysis. 
The underlined scale is closely related to the presence of a material content, which reflects the energy scale involved.

Without matter, it is difficult to define a scale. 
Therefore, an empty universe would be expected to be scale invariant. 
This is confirmed by the scale invariance of Maxwell's equations in vacuum, 
which are the equations that govern the dynamics of the electromagnetic fields. 
The field equations of general relativity are also scale invariant for empty space with zero cosmological constant. 
However, it is still an open question how much matter is needed to break scale invariance. 
This question is particularly relevant to cosmology and the evolution of the universe.

\subsection{Einstein General Relativity and Weyl Integrable Geometry}

Albert Einstein's General Theory of Relativity (EGR) is based on the premise that 
a torsion-free covariant connection is metric-compatible and ensures the 
preservation of the length of vectors along geodesics. 
The theory has been successfully tested at a variety of scales, 
starting with local Earth laboratories, the Solar System, galactic scales via light-bending effects, 
and even on an extragalactic level via the observation of gravitational waves. 
EGR is also the foundation of modern cosmology and astrophysics. 
However, at galactic and cosmic scales, some new and mysterious phenomena have emerged. 
These phenomena are often attributed to unknown matter particles or fields that 
have yet to be detected in our laboratories, 
hence the suggestive names "dark matter" and "dark energy."

Since no new particles or fields have been detected in Earth labs for more than twenty years, 
it is reasonable to revisit some old ideas that have been proposed as modifications of Einstein's general relativity. 
In 1918, Weyl proposed an extension by adding local gauge (scale) invariance \cite{Weyl23}. 
Other approaches were more radical, such as Kaluza–Klein unification theory, which adds extra dimensions. 
The return to the usual 4D spacetime could be done using projective relativity theory via Jordan conformal equivalence, 
but with at least one additional scalar field. 
Such theories are also known as Jordan–Brans–Dicke 
scalar-tensor gravitation theories \cite{Brans14,Faraoni+99,2002PhLB..530...20B}. 
In most such theories, there is a major drawback: a varying Newton constant $G$. 
Some theories go even further to consider spatially varying $G$ gravity \cite{2023MNRAS.519.1277C}. 
No such variations have been observed yet, so we prefer to view Newton's gravitational constant $G$ as constant 
despite some current experimental issues \cite{Xue20}.

The above discussion raises the question of whether the mysterious "dark" phenomena could be artifacts of 
non-zero change in of the length of vectors along geodesics, 
which is often negligible and has almost zero value $(\delta\left\Vert\overrightarrow{v}\right\Vert \approx0)$, 
but could accumulate over cosmic distances and fool us into thinking that 
the observed phenomena may be due to dark matter and/or dark energy. 
Weyl proposed an extension of Einstein's General Relativity as soon as EGR was proposed, 
with the desired properties of local gauge (scale) invariance. 
This has the consequence that lengths may not be preserved upon parallel transport. 
However, it was quickly argued that such a model would result in a path-dependent phenomenon, 
and thus contradict observations. 
A remedy was found to this objection by introducing Weyl Integrable Geometry (WIG) \cite{Weyl23}, 
where the lengths of vectors are conserved only along closed paths 
($\varoint\delta\left\Vert\overrightarrow{v}\right\Vert =0$).

This concept leads to scale-invariant cosmology, as proposed by Dirac and Canuto in \cite{Dirac73, Canuto77}. 
This formulation of Weyl's original idea overcomes Einstein's objection! Furthermore, 
given that all we can observe about the distant universe are waves that reach us, 
the condition for Weyl integrable geometry is essentially stating that the information that 
reaches us via different paths constructively interferes to create a consistent image of the source.

One approach to building a WIG model is to consider a conformal transformation of the metric field 
$g'_{\mu\nu}=\lambda^2g_{\mu\nu}$ and apply it to various observational phenomena. 
As we will see in the discussion below, 
the requirement for homogeneous and isotropic space restricts the field $\lambda$ 
to depend only on cosmic time and not on spatial coordinates. 
The weak field limit of such a WIG model results in an additional acceleration in the equation of motion that 
is proportional to the particle's velocity.

Such behavior is somewhat similar to that of Jordan–Brans–Dicke scalar-tensor gravitation, 
but the conformal factor $\lambda$ does not appear to be 
a typical scalar field as in the Jordan–Brans–Dicke theory \cite{Brans14,Faraoni+99}.
Furthermore, the Scale Invariant Vacuum (SIV) concept provides a way to 
determine the specific functional form of  $\lambda(t)$ 
as it applies to FLRW cosmology and its WIG extension.

The functional form of  $\lambda$ leads to a specific functional form 
for the rate of change of its logarithm $\kappa=-d(\ln{\lambda})/dt$, 
which controls the strength of an additional SIV acceleration $\kappa\,v$
in the SIV modified equations of motion.
It is important to note that the additional acceleration in the equations of motion, 
which is proportional to the velocity of a particle, can also be justified by requiring reparametrization symmetry. 
Reparametrization invariance is often overlooked as being part of the general covariance that 
guarantees that physics is independent of the observer's coordinate system. 
However, reparametrization symmetry is much more than that; 
it is about the physics being independent of the choice of parametrization of a process under study. 
Not implementing reparametrization invariance in a model could lead to 
un-proper time parametrization,\footnote{ The proper time parametrization of a process 
is the time, up to a constant scale factor, measured by a standard clock in the rest/co-moving inertial frame for the process.
Any other time parametrization will be considered un-proper time parametrization since it is arbitrary.
In this respect the coordinate time of any arbitrary coordinate system will provide an example of  
un-proper time parametrization for the process when described in that coordinate frame.} 
which seems to induce "fictitious forces" in the equations of motion, 
similar to the forces derived in the weak field SIV regime \cite{sym13030379}. 
This is a puzzling observation that may help us better understand nature given its relation to 
some of the key properties of physical systems \cite{2021Symm...13..522G}.

\section{Mathematical Framework}

The  framework for the Scale Invariant Vacuum paradigm is based on the 
Weyl Integrable Geometry and the Dirac co-calculus as mathematical tools for 
description of nature \cite{Weyl23,Dirac73}.
For a more modern treatment of the scale invariant gravity idea see \cite{2018GReGr..50...80W},
that is based on the Cartan's formalism and along the more traditional scalar field approach, 
which due to its abstractness seems to have stayed disconnected from observational tests,
apart of a few papers on the model parameters for conformal cosmology 
\cite{2010IJMPD..19.1875Z,2017PPNL...14..368P} where dark matter and energy 
seem to be replaced by the concept of rigid matter, which is still observationally questionable as its dark counterparts.
Here, our approach is more traditional, physically motivated and with as little general abstraction as possible.
For more mathematical details we refer the reader to the companion paper on the
``Action Principle for Scale Invariance and Applications (Part I)'' \cite{AM&VG'2310}.

\subsection{Weyl Integrable Geometry and Dirac Co-Calculus}

The original Weyl geometry uses a metric tensor field $g_{\mu\nu}$, 
along with a ``connexion'' vector field $\kappa_{\mu}$, and a scalar field $\lambda$.
Here we use the french spelling of the word connection to avoid misinterpretation and
confusion with the usual meaning and use of a connection vector field.
In the Weyl Integrable Geometry (WIG), the ``connexion'' vector field $\kappa_{\mu}$ 
is not an independent  field, but it is derivable from the scalar field $\lambda$.

\begin{equation}
\kappa_{\mu}=-\partial_{\mu}\ln(\lambda)
\label{connexion}
\end{equation}

This form of the ``connexion'' vector field $\kappa_{\mu}$ guarantees its irrelevance, 
in the covariant derivatives, upon integration over closed paths.
That is, $\varoint \kappa_{\mu}dx^{\mu} =0$. In other words, $\kappa_{\mu}dx^{\mu}$ represents a closed 1-form; furthermore, 
it is an exact form, as \eqref{connexion} implies 
$\kappa_{\mu}dx^{\mu}=-d\ln{\lambda}$.
Thus, the scalar function $\lambda$ plays a key role in the Weyl Integrable Geometry.
Its physical meaning is related to the freedom 
of choice of a local scale gauge. Thus,  $\lambda$ relates to 
the changes in the equations of a physical system upon change in scale 
via local re-scaling $l'\rightarrow\lambda(x)l$. 
Such change could be induced via a local conformal transformation of the coordinates,
in which case it is part of the general diffeomorphism symmetry,
or it could be only a metric conformal transformation without any associated coordinate transformation.

\subsubsection{Gauge Change and (co-) covariant Derivatives}
The covariant derivatives utilize the rules of the Dirac co-calculus  \cite{Dirac73} 
where tensors also have co-tensor powers based on the way they transform upon change of scale.
For the metric tensor $g_{\mu\nu}$ this power is $\Pi(g_{\mu\nu})=2$. 
This follows from  the way the length of a line segment $ds$ 
is defined via the usual expression $ds^2=g_{\mu\nu}dx^{\mu}dx^{\nu}$.

$$l'\rightarrow\lambda(x)l\Leftrightarrow ds'=\lambda ds \Rightarrow g'_{\mu\nu}=\lambda^{2}g_{\mu\nu}.$$

Thus, $g^{\mu\nu}$ is having co-tensor power of $\Pi(g^{\mu\nu})=-2$ in order to make the 
Kronecker $\delta$ a scale invariant object ($g_{\mu\nu}g^{\nu\rho}=\delta_{\mu}^{\rho}$).
That is, a co-tensor is of power $n$ when, upon local scale change, it satisfies:

\begin{equation}
l'\rightarrow\lambda(x)l :\;  Y'_{\mu\nu}\rightarrow\lambda^{n}Y_{\mu\nu}
\label{co-tensor}
\end{equation}

\subsubsection{Dirac Co-Calculus}
In the Dirac co-calculus, this results in the appearance of the ``connexion'' vector field $\kappa_{\mu}$ 
in the  covariant derivatives of scalars, vectors, and tensors (see Table \ref{Table1}):

\begin{table}[h]
\begin{center}
\begin{tabular}{cc} 
\hline
\textbf{Co-Tensor Type} & \textbf{Mathematical Expression} \\
\hline
co-scalar & $S_{*\mu}=\partial_{\mu}S-n\kappa_{\mu}S$,\\
co-vector & $A_{\nu*\mu}=\partial_{\mu}A_{\nu}-\;^{*}\Gamma_{\nu\mu}^{\alpha}A_{\alpha}-n\kappa_{\nu}A_{\mu}$,\\
co-covector & $A_{*\mu}^{\nu}=\partial_{\mu}A^{\nu}+\;^{*}\Gamma_{\mu\alpha}^{\nu}A^{\alpha}-nk^{\nu}A_{\mu}$.\\
\hline
\end{tabular}
\end{center}
\caption{\small Derivatives for co-tensors of power $n$.}\label{Table1} 
\end{table}%

\noindent 
where the usual Christoffel symbol $\Gamma_{\mu\alpha}^{\nu}$ is replaced by
\begin{equation}
^{*}\Gamma_{\mu\alpha}^{\nu}=\Gamma_{\mu\alpha}^{\nu}
+g_{\mu\alpha}k^{\nu}-g_{\mu}^{\nu}\kappa_{\alpha}-g_{\alpha}^{\nu}\kappa_{\mu}.
\label{eq:Christoffel*}
\end{equation}

The corresponding equation of the geodesics within the WIG 
was first introduced in 1973 by \cite{Dirac73} 
and in the weak-field limit was re-derived in 1979 by \cite{MBouvier79}
($u^{\mu}=dx^{\mu}/{ds}$ is the four-velocity):
\begin{equation}
u_{*\nu}^{\mu}=0\Rightarrow\frac{du^{\mu}}{ds}+\,^{*}\Gamma_{\nu\rho}^{\mu}u^{\nu}u^{\rho}
+\kappa_{\nu}u^{\nu}u^{\mu}=0\,.
\label{eq:geodesics*}
\end{equation}

This geodesic equation has also been derived from 
reparametrization-invariant action in 1978 by \cite{BouvM78}:
\begin{equation*}
\delta\mathcal{A}=\intop_{P_{0}}^{P_{1}}\delta\left(d\widetilde{s}\right)
=\int\delta\left(\beta ds\right)=\int\delta\left(\beta\frac{ds}{d\tau}\right)d\tau=0.
\label{eq:action}
\end{equation*}

\subsection{Consequences of going beyond the EGR}
\label{sec:beyondEGR}

Before we go into the specific examples, 
such as FLRW cosmology and weak-field limit, there are some remarks to be made.
By using \eqref{eq:Christoffel*} in \eqref{eq:geodesics*}, one can see that 
the usual EGR equations of motion receive extra terms proportional to the four-velocity 
and \mbox{its normalization}:

\begin{equation}
\frac{du^{\mu}}{ds}+\Gamma_{\nu\rho}^{\mu}u^{\nu}u^{\rho}
=(\kappa\cdot u)u^{\mu}-(u\cdot u)\kappa^{\mu}
\label{eq:geodesics+}
\end{equation}

In the weak-field approximation within the SIV, one assumes an isotropic and homogeneous space
for the explicit derivation of the new terms beyond the usual Newtonian equations \cite{BouvM78}. 
As seen from \eqref{eq:geodesics+}, the result is a velocity dependent extra term  
$\kappa_0\vec{v}$ with $\kappa_0=-\dot{\lambda}/\lambda$,
while the special components are set to zero ($\kappa_i=0,\;i=1,2,3$)
due to the assumption of isotropic and homogeneous space. 
At this point, it is important to stress that the usual normalization 
for the four-velocity, $u\cdot u=\pm1$ with sign 
related to the signature of the metric tensor $g_{\mu\nu}$,
is a special choice of parametrization---the proper-time
parametrization $\tau$.
We denote a general parametrization in \eqref{eq:geodesics+} 
with $s$, while $\tau$ is reserved for the proper time, 
and $t$ is the coordinate time parametrization.

Similar extra term ($\kappa_0\vec{v}$) was recently obtained \cite{sym13030379} 
as a consequence of reparametrization invariant mathematical modeling
but without the need for a weak-field approximation. 
That is, insisting on reparametrization symmetry for the equations of motion
demands such term to be present in order to account for the change of
parametrization within a chosen coordinate system.
Within the proper time-parametrization one usually has $\kappa_0=0$.
However, if one assumes that the equations used for the process under study are 
parametrized via the proper time-parametrization but 
relies on the observer coordinate time,
without including the appropriate $\kappa$-term then one has 
incorrect modeling with un-proper time parametrization instead
because coordinate time is often quite different from the proper time of a process.
{\it Therefore, not accounting for reparametrization symmetry leads to missing terms 
in the mathematical formulas utilized in the modeling of a system.}
The $\kappa$-term is required by reparametrization symmetry. When properly accounted for, 
it appears as a velocity-dependent fictitious acceleration \cite{sym13030379}. 
The term $\kappa_0 \vec{v}$ is necessary to restore the broken symmetry, 
which is the reparametrization invariance of the process under consideration. 
To demonstrate this, we can apply an arbitrary time reparameterization $\lambda = dt/ds$. 
Then, the first term on the left-hand side of equation \eqref{eq:geodesics+} becomes:

\begin{equation}
\lambda \frac {d }{dt} \left(\lambda \frac {d\vec{r}}{dt} \right)\, 
= \lambda^2 \frac {d^2\vec{r}}{dt^2}\,+ \lambda\dot{\lambda}\frac{d\vec{r}}{dt} .
\label{eq:re-parametrization}
\end{equation}
\noindent
By moving the velocity-linear term to the right-hand side of \eqref{eq:geodesics+}, 
using $\kappa(t) = -\dot{\lambda}/\lambda$ after dividing by $\lambda^2$, 
we obtain a $\kappa_0 \vec{v}$-like term on the right-hand side. 
If we perform this manipulation in the absence of the $\kappa_0 \vec{v}$ term 
on the left-hand side of \eqref{eq:geodesics+}, the term is generated. 
If the $\tilde\kappa$ term is present in the equations, 
it is transformed into $\tilde\kappa \rightarrow \kappa + \tilde\kappa$.

Unlike in SIV, where the time reparameterization $\lambda(t)$ is typically justified to be $t_0/t$, 
for reparameterization symmetry, the time dependence of $\lambda(t)$ can be arbitrary. 
As discussed in \cite{sym13030379}, the extra term $\kappa_0 \vec{v}$ is not expected to be present 
when the time parametrization of the process is the proper time of the system. 
{\it Thus, a term of the form $\kappa \vec{v}$ can be viewed as necessary 
for restoration of the re-parametrization symmetry and an indication of 
un-proper time parametrization of a process under consideration when omitted.}

In the context of FLRW cosmology, under the assumption of homogeneity and isotropy of space, 
the proper time parameterization is given by 
$-c^2d\tau^2 = -c^2dt^2 + a(t)^2d\Sigma^2$, where $c$ is the speed of light (set to 1), 
$\Sigma$ is a 3D space of uniform curvature, and $a(t)$ is the scale factor for 3D space. 
Here, $\tau$ is the proper time of the cosmological evolution, 
while $t$ is the coordinate time of an observer studying the cosmic evolution.

Upon transitioning to WIG, one introduces a multiplicative conformal factor $\lambda(x)$. 
In the case of $\lambda(t)$ (time dependence only), 
one can argue that this factor can be absorbed into $a(t)$ by redefining the coordinate time $t$ via $d\tilde{t} = \lambda(t) dt$. 
However, this does not guarantee proper time parameterization in general. 
Therefore, it is likely that the FLRW cosmology equations with missing velocity-dependent terms 
will have improper time parameterization, unless one ensures that reparameterization symmetry is restored.

\subsection{Scale Invariant Cosmology}

The scale invariant cosmology equations, first introduced by Dirac in 1973 \cite{Dirac73} and 
re-derived by Canuto in 1977 \cite{Canuto77}, are based on the corresponding expressions of 
the Ricci tensor and a relevant extension of the Einstein equations. We have recently  revisited the 
topic to bring it into focus and aligned the SIV paradigm \cite{AM&VG'2310}. In what follows,
we paint a broad stroke picture of the equations and their consequences.

\subsubsection{The Einstein Equation for Weyl's Geometry}

Upon the metic conformal transformation $g'_{\mu\nu} = \lambda^2 g_{\mu\nu}$ 
from Weyl's framework to the EGR framework, 
where $g'_{\mu\nu}$ is the metric tensor in the  EGR framework, 
a simple relation is induced between the Ricci tensor and scalar within 
Weyl's Integrable Geometry and the Einstein framework. 
In our convention, a prime ($\prime$) is used to denote EGR framework objects:

\begin{eqnarray*}
R_{\mu\nu}=R'_{\mu\nu}-\kappa_{\mu;\nu}-\kappa_{\nu;\mu}-2\kappa_{\mu}\kappa_{\nu}
+2g_{\mu\nu}\kappa^{\alpha}\kappa_{\alpha}-g_{\mu\nu}\kappa_{;\alpha}^{\alpha}\,,\\
R=R'+6\kappa^{\alpha}\kappa_{\alpha}-6\kappa_{;\alpha}^{\alpha}\,.
\end{eqnarray*}
By using these expressions, we can extent the standard EGR equation into: 
\begin{eqnarray}
R_{\mu\nu}-\frac{1}{2}\ g_{\mu\nu}R=-8\pi GT_{\mu\nu}-\Lambda\,g_{\mu\nu}\,,\\
R'_{\mu\nu}-\frac{1}{2}\ g_{\mu\nu}R'-\kappa_{\mu;\nu}-\kappa_{\nu;\mu}
-2\kappa_{\mu}\kappa_{\nu}+2g_{\mu\nu}\kappa_{;\alpha}^{\alpha}-g_{\mu\nu}\kappa^{\alpha}\kappa_{\alpha}=\nonumber \\
-8\pi GT_{\mu\nu}-\Lambda\,g_{\mu\nu}\,.
\label{field}
\end{eqnarray}

Here $\Lambda$ is in WIG and is expected that $\Lambda=\lambda^{2}\Lambda_{\mathrm{E}}$,
with $\Lambda_{\mathrm{E}}$ beeing the Einstein cosmological constant in EGR.
This relationship guarantees the explicit scale invariance of the equations.
This makes explicit the appearance of $\Lambda_{\mathrm{E}}$ as invariant scalar (in-scalar), since then:
$\Lambda\,g_{\mu\nu}=\lambda^{2}\Lambda_{\mathrm{E}}\,g_{\mu\nu}=\Lambda_{\mathrm{E}}\,g'_{\mu\nu}$.
That is, the co-scaler power of $\Lambda$ in WIG is $\Pi(\Lambda)=-2$.

The equations above are a generalization of the original equations due to Einstein. 
As a result, there is a larger class of gauge symmetries, which must be fixed by an appropriate gauge choice
in order to do practical studies. Dirac considered the large numbers hypothesis for his gauge choice \cite{Dirac74} . 
Below, we consider a different gauge fixing. 

The scale-invariant FLRW cosmology equations were first introduced 
in 1977 by  \cite{Canuto77} in the following form:
\begin{eqnarray}
\frac{8\,\pi G\varrho}{3}=\frac{k}{a^{2}}+\frac{\dot{a}^{2}}{a^{2}}+2\,\frac{\dot{\lambda}\,\dot{a}}{\lambda\,a}
+\frac{\dot{\lambda}^{2}}{\lambda^{2}}-\frac{\Lambda_{\mathrm{E}}\lambda^{2}}{3}\,,\label{E1p}\\
-8\,\pi Gp=\frac{k}{a^{2}}+2\frac{\ddot{a}}{a}+2\frac{\ddot{\lambda}}{\lambda}+\frac{\dot{a}^{2}}{a^{2}}
+4\frac{\dot{a}\,\dot{\lambda}}{a\,\lambda}-\frac{\dot{\lambda^{2}}}{\lambda^{2}}-\Lambda_{\mathrm{E}}\,\lambda^{2}\,.\label{E2p}
\end{eqnarray}

The above equations reduce to the standard FLRW equations in the limit $\lambda=const=1$. 
The scaling of $\Lambda$ with $\lambda$ has been utilized to revisit the cosmological constant problem 
within quantum cosmology \cite{GueorM20}; this resulted in the conclusion that our Universe is unusually large, 
given that the mean size of all universes, where EGR holds, was calculated to be of the order of the Planck scale. 
In that study, $\lambda=const$ was a key assumption, as the various universes were expected to 
obey the  EGR equations.\footnote{
The question: "what would be the expected mean size of a universe, if the condition  
$\lambda=const$ is relaxed, remains an open question for an ensemble of WIG-universes"
was already stressed out in our previous review paper \cite{2022Univ....8..213G}.}

\subsubsection{The Scale Invariant Vacuum Gauge at $T=0$ and $R'=0$} 

The idea of the Scale Invariant Vacuum was introduced first in 2017 by \cite{Maeder17a}.
For an empty universe model, the de Sitter metric is conformal to the Minkowski metric,
thus, $R'_{\mu \nu}$  is vanishing \cite{Maeder17a}. 
Therefore, for conformally flat metric, 
that is, Ricci flat ($R'_{\mu\nu}=0$) Einstein vacuum ($T_{\mu\nu}=0$),
the following vacuum equation can be obtained using \eqref{field}:
\begin{equation}
\kappa_{\mu;\nu}+\kappa_{\nu;\mu}+2\kappa_{\mu}\kappa_{\nu}
-2g_{\mu\nu}\kappa_{;\alpha}^{\alpha}+g_{\mu\nu}\kappa^{\alpha}\kappa_{\alpha}=\Lambda\,g_{\mu\nu}\label{SIV}
\end{equation}

\noindent
For homogeneous and isotropic space ($\partial_{i}\lambda=0$), 
only $\kappa_{0}=-\dot{\lambda}/\lambda$ and its time
derivative $\dot{\kappa}_{0}=-\kappa_{0}^{2}$ can be non-zero.
As a corollary of \eqref{SIV}, one can derive the following set of equations \cite{Maeder17a}:
\begin{eqnarray}
\ 3\,\frac{\dot{\lambda}^{2}}{\lambda^{2}}\,=\Lambda\,,\quad\mathrm{and}\quad2\frac{\ddot{\lambda}}{\lambda}
-\frac{\dot{\lambda}^{2}}{\lambda^{2}}\,=\Lambda\,,\label{SIV1}\\
\mathrm{or}\quad\frac{\ddot{\lambda}}{\lambda}\,=\,2\,\frac{\dot{\lambda}^{2}}{\lambda^{2}}\,,
\quad\mathrm{and}\quad\frac{\ddot{\lambda}}{\lambda}-\frac{\dot{\lambda}^{2}}{\lambda^{2}}\,=\frac{\Lambda}{3}\,.\label{SIV2}
\end{eqnarray}

\noindent
These equations can be derived by using the time and space components of the equations, 
or by looking at the relevant trace invariant along with the relationship $\dot{\kappa}_{0}=-\kappa_{0}^{2}$.
Any one pair among these equations is sufficient to prove the validity of the other pair of equations. 

\begin{thm}
Using the SIV Equations \eqref{SIV1} or \eqref{SIV2} with $\Lambda=\lambda^{2}\Lambda_{E}$ one has: 
\begin{equation}
\Lambda_{E}=3\frac{\dot{\lambda^{2}}}{\lambda^{4}},\quad\mathrm{with}\quad\frac{d\Lambda_{E}}{dt}=0.
\label{SIV-gauge}
\end{equation}
\end{thm}
 
\begin{cor}
The solution of the SIV gauge equations is then:
\begin{equation} 
\lambda=t_{0}/t, 
\label{lambda(t)} 
\end{equation}
with $t_0=\sqrt{3/(c^2\Lambda_{E})}$ where $c$ is the speed of light usually set to 1.
\end{cor} 

The choice of such gauge for $\lambda$ can be used to replace the Dirac's  large numbers hypothesis invoked by 
\cite{Canuto77}. This is what we refer to as a Scale Invariant Vacuum (SIV) gauge for $\lambda$.

Even more, now we can have an alternative viewpoint on \eqref{field} and \eqref{SIV}.
Since \eqref{field}  is scale invariant then one does not have to consider zero case for 
$T_{\mu\nu}$ and $R_{\mu\nu}$ in general, but if the scale factor $\lambda$ satisfies \eqref{SIV},
then all the $\kappa$ terms and the $\Lambda$ term in \eqref{field} will cancel out 
leaving  us with the standard EGR equation with zero cosmological constant.
{\it Thus, a proper choice of $\lambda$ gauge satisfying \eqref{SIV}
results in the standard Einstein equation with no cosmological constant!}
This is easily seen in the case of homogeneous and isotropic universe or 
when requiring only reparametrization invariance, 
then both cases are resulting in \eqref{SIV1} and \eqref{SIV2} along with \eqref{SIV-gauge}.
{\it If one takes the reparametrization symmetry viewpoint then the presence of a non-zero 
cosmological constant is indication of un-proper time parametrization that can be cured
upon suitable new time gauge deduced by the appropriate choice of $\lambda$.}

Upon the use of the SIV gauge, first in  2017 by \cite{Maeder17a}, one observes that 
{\it the cosmological constant disappears} from Equations \eqref{E1p} and \eqref{E2p}:  
\begin{eqnarray}
\frac{8\,\pi G\varrho}{3}=\frac{k}{a^{2}}+\frac{\dot{a}^{2}}{a^{2}}+2\,\frac{\dot{a}\dot{\lambda}}{a\lambda}\,,\label{E1}\\
-8\,\pi Gp=\frac{k}{a^{2}}+2\frac{\ddot{a}}{a}+\frac{\dot{a^{2}}}{a^{2}}+4\frac{\dot{a}\dot{\lambda}}{a\lambda}\,.\label{E2}
\end{eqnarray}

The solutions of these equations have been discussed in details in \cite{Maeder17a}, together 
 with various cosmological properties concerning the Hubble-Lema\^{i}tre and deceleration  parameters, the 
 cosmological distances and  different
 cosmological tests. The redshift drifts appear as one of the most promising cosmological tests  \cite{MaedGueor20a}.
 Here, we limit the discussion to a few points pertinent to the subject of the paper.
Analytical solutions  for the flat SIV models with $k=0$  have been found 
for the  matter  \cite{Jesus18} and radiation  \cite{Maeder19} dominated  models. 
In the former case, we have a simple expression:
\begin{equation}
a(t) \, = \, \left[\frac{t^3 -\Omega_{\mathrm{m}}}{1 - \Omega_{\mathrm{m}}} \right]^{2/3}\, .
\label{Jesus}
\end{equation}
\noindent
It is expressed in the SIV-timescale $t$ where at present  $t_0=1$ and  $a(t_0)=1$. 
Such solutions are illustrated in Fig. \ref{rates}.
They are lying relatively close to the $\Lambda$CDM ones, the differences being larger for lower $\Omega_{\mathrm{m}}$.
This is a general property: {\it{the effects of scale invariance are always larger for the lower matter densities, being the
largest ones for the empty space. }}
As usual, here $\Omega_{\mathrm{m}}= \varrho/\varrho_{\mathrm{c}}$  with $\varrho_{\mathrm{c}}=3H^2_0/(8\pi G)$.
Remarkably, Eqs. (\ref{E1}) and (\ref{E2}) allow flatness for different values of  $\Omega_{\mathrm{m}}$. 
It follows from  \eqref{Jesus} that the initial time at $a( t_{\mathrm{in}})=0$ is related to the value of $\Omega_\mathrm{m}$:

\begin{equation}
 t_{\mathrm{in}} \, = \, \Omega^{1/3}_{\mathrm{m}}\,.
 \label{tin}
 \end{equation}
\noindent
The Hubble parameter and $\kappa_0(t)=-\dot{\lambda}/\lambda$ are then, 
in the  timescale $t$ (which goes from $t_{\mathrm{in}}$ at the Big-Bang to $t_0=1$ at present):
 \begin{equation}
 H(t) =\frac{\dot{a}}{a} =\frac{2 \, t^2}{t^3- \Omega_{\mathrm{m}}}\,,\;
 \text{and}\;\kappa_0(t)=-\frac{\dot{\lambda}}{\lambda}=\frac{1}{t}.
 \label{HM}
 \end{equation}
From Eqs. (\ref{Jesus}) and (\ref{HM}), we  see that there is no meaningful scale invariant solution
for an expanding Universe ($H>0$) with $\Omega_{\mathrm{m}}$ equal or larger than 1.
Thus, the model solutions   are  quite  consistent with  the  causality relations  discussed by \cite{SIV-Inflation'21}.

The usual timescale $\tau$  in years or seconds  is $\tau_0= 13.8$ Gyr at  present  \cite{Frie08} 
and  $\tau_{\mathrm{in }}=0$ at the Big-Bang. One can change from the SIV-time $t$ to the usual time scale $\tau$
by using the relationship ansatz  \cite{2022arXiv220413560M}:
\begin{equation}
\frac{\tau - \tau_{\mathrm{in}}}{\tau_0 - \tau_{\mathrm{in}}} = \frac{t - t_{\mathrm{in}}}{t_0 - t_{\mathrm{in}}}\, ,
\end{equation}
which is expressing that the age fraction with respect to the present age is the same in both timescales.
This ansatz gives: 
\begin{equation}
\tau \,= \, \tau_0 \, \frac{t- \Omega^{1/3}_{\mathrm{m}}}{1- \Omega^{1/3}_{\mathrm{m}}} \,  \quad \mathrm{and} \; \;
  t \,= \, \Omega^{1/3}_{\mathrm{m}} + \frac{\tau}{\tau_0} (1- \Omega^{1/3}_{\mathrm{m}}) \,,
\label{T2}
\end{equation}
The relevant derivatives are constants depending on $t_{\mathrm{in}}=\Omega^{1/3}_{\mathrm{m}}$ and  $\tau_0$ only:
\begin{equation}
\frac{d\tau}{dt} \, = \, \frac{\tau_0}{1-\Omega^{1/3}_{\mathrm{m}}}\,, \quad \mathrm{and}\; \;
\frac{dt}{d\tau} \, = \, \frac{1-\Omega^{1/3}_{\mathrm{m}}}{\tau_0}\,.
\label{dT1}
\end{equation}
For larger  $\Omega_{\mathrm{m}}$,   timescale $t$ is squeezed over a smaller fraction of the interval
0 to 1, (which reduces the range of $\lambda$ over the ages).
Using the above expressions one can write the Hubble parameter in the  usual time scale $\tau$
via its expression in the $t$-scale:
\begin{equation}
H(\tau) =\frac{\dot{a}}{a} =H(t)\frac{dt}{d\tau}=H(t)\frac{1-\Omega^{1/3}_{\mathrm{m}}}{\tau_0}.
\label{H(tau)}
\end{equation}
This finally gives for the Hubble constant:
\begin{equation}
H_0 =\frac{2}{1-\Omega_{\mathrm{m}}}\frac{1-\Omega^{1/3}_{\mathrm{m}}}{\tau_0}.
\label{H0}
\end{equation}
The last factor could be recognized as $\kappa_0(\tau_0)$. 
To see this one can utilize the equations \eqref{T2} and  \eqref{dT1} 
to switch from the SIV-time $t$ to the conventional time $\tau$ scale \cite{2022arXiv220413560M}
in order to obtain:
\begin{eqnarray}
\kappa_0(\tau) =-\frac{\dot{\lambda}}{\lambda}=\kappa_0(t)\frac{dt}{d\tau}=\frac{1-t_\text{in}}{t\,\tau_0}
=\frac{1-t_\text{in}}{\tau_0}\frac{1}{t_\text{in}+(1-t_\text{in})(\tau/\tau_0)}=\frac{\psi(\tau)}{\tau_0}\label{psi(tau)}\,,\\
\Rightarrow\kappa_0(\tau_0)=\frac{1-\Omega^{1/3}_{\mathrm{m}}}{\tau_0}\; \text{and}\; 
\psi(\tau)=\frac{1-t_\text{in}}{t_\text{in}+(1-t_\text{in})(\tau/\tau_0)}\,.
\label{kappa00}
\end{eqnarray}

\section{Comparisons and Applications}

Over the past few years the authors have published a series of papers that compare 
the predictions and outcomes of the SIV paradigm to observations. 
The main results and outcomes are discussed in this section.

\subsection{Scale Factor $a(t)$ within $\Lambda$CDM and SIV \cite{Maeder17a} }
\begin{figure}[h]
\centering 
\includegraphics[width=0.8\textwidth]{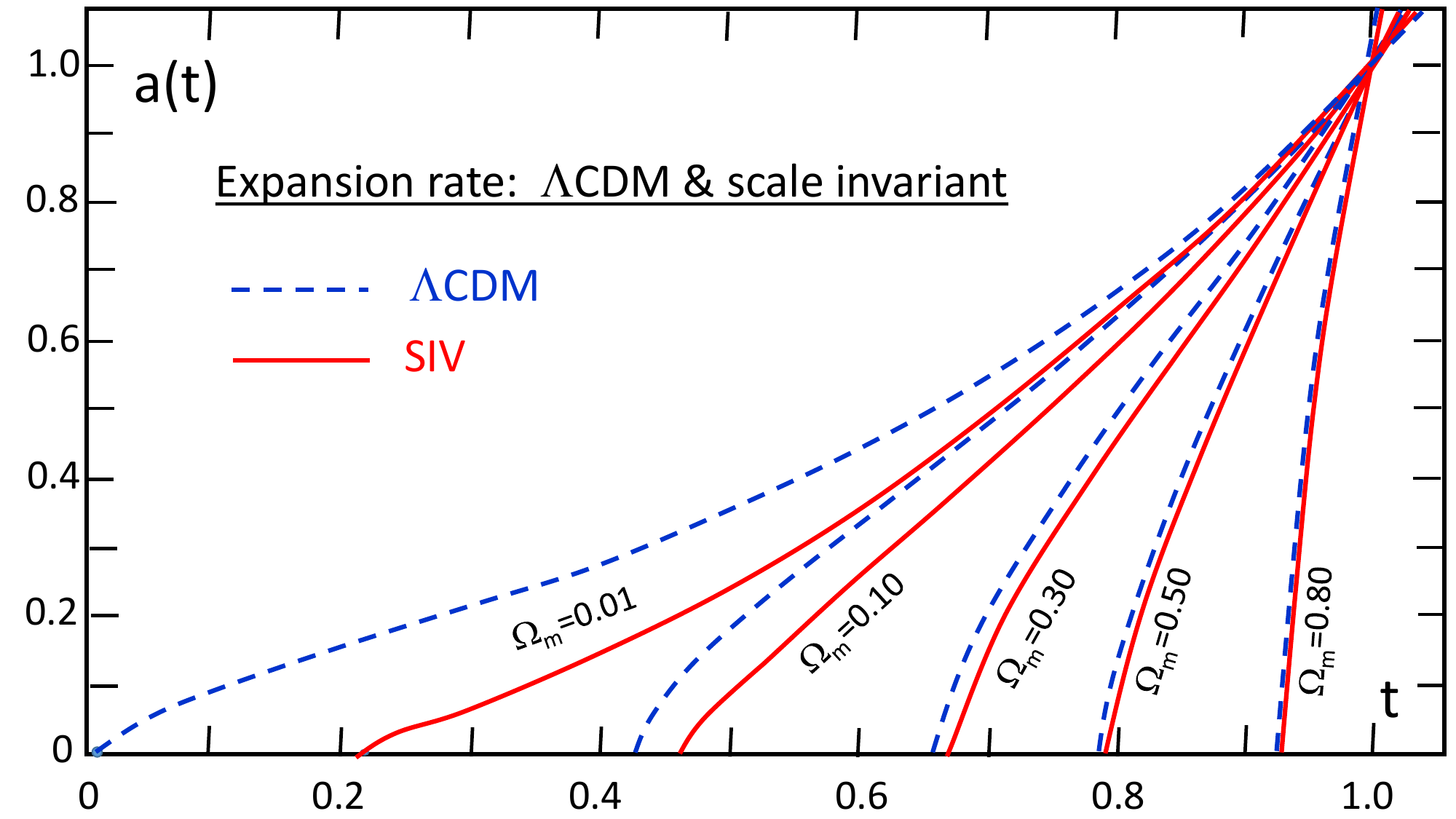} 
\caption{\small 
Expansion rates $a(t)$ as a function of time $t$ in the flat
($k=0$) $\Lambda$CDM and SIV models in the matter dominated era.
The curves are labeled by the values of $\Omega_{\mathrm{m}}$;
here $\Omega_{\mathrm{m}}= \varrho/\varrho_{\mathrm{c}}$  with $\varrho_{\mathrm{c}}=3H^2_0/(8\pi G)$.
Drawing originally published in \cite{Maeder17a}.}
\label{rates} 
\end{figure}

The implications of the Scale Invariant Vacuum Paradigm for Cosmology were 
first discussed by \cite{Maeder17a}  and later reviewed by \cite{MaedGueor20a}. 
In this paper, we use the SIV equations \eqref{E1} and \eqref{E2},  along with the gauge fixing \eqref{SIV-gauge},  
which implies  \eqref{lambda(t)}, that is, $\lambda=t_0/t$ with $t_0$ indicating 
the current age of the Universe since the Big Bang defined as $a(t_\text{in})=0$ at some past moment $t_\text{in}$.

The most important point in comparing $\Lambda$CDM and SIV cosmology models is the existence of 
SIV cosmology with slightly different parameters but almost the same 
curve for the standard scale parameter $a(t)$ 
when the time scale is set so that  $t_0=1$ at the present epoch 
\cite{Maeder17a,MaedGueor20a}. 

As seen in Figure~\ref{rates}, the differences between the $\Lambda$CDM and SIV models 
decline for increasing matter densities \cite{Maeder17a}.
Furthermore, the SIV solutions are lying relatively close to the $\Lambda$CDM ones, 
the differences being larger for lower $\Omega_{\mathrm{m}}$.
This is a general property: {\it{the effects of scale invariance are always larger for the lower matter densities, 
being the largest when approaching the empty space. }}

%
%
%

\subsection{Possible differentiators of SIV from $\Lambda$CDM \cite{MaedGueor20a}}

The major property of SIV cosmology is that it naturally predicts an acceleration of the expansion. This is the consequence 
of the additional term in Eqs. (\ref{E1}) and (\ref{E2}) which predicts an acceleration of the motion in the direction of the velocity.
If the Universe were to contract, it would also receive an additional acceleration favoring a contraction.

Several observational tests of the SIV cosmology are performed and discussed in details in \cite{MaedGueor20a}. 
For example, based on Fig. 3 in  \cite{MaedGueor20a} one can see
that the relation between the Hubble constant $H_0$ and the age of the Universe in the SIV Cosmology
is suggesting a range of values for $\Omega_m$ between 0.15 and 0.25 depending on the choice for 
$H_0$ using either the distance ladder or Planck collaboration measurements.

\begin{figure}[h]
\begin{center}
\includegraphics[width=11.5cm,height=8.0cm]{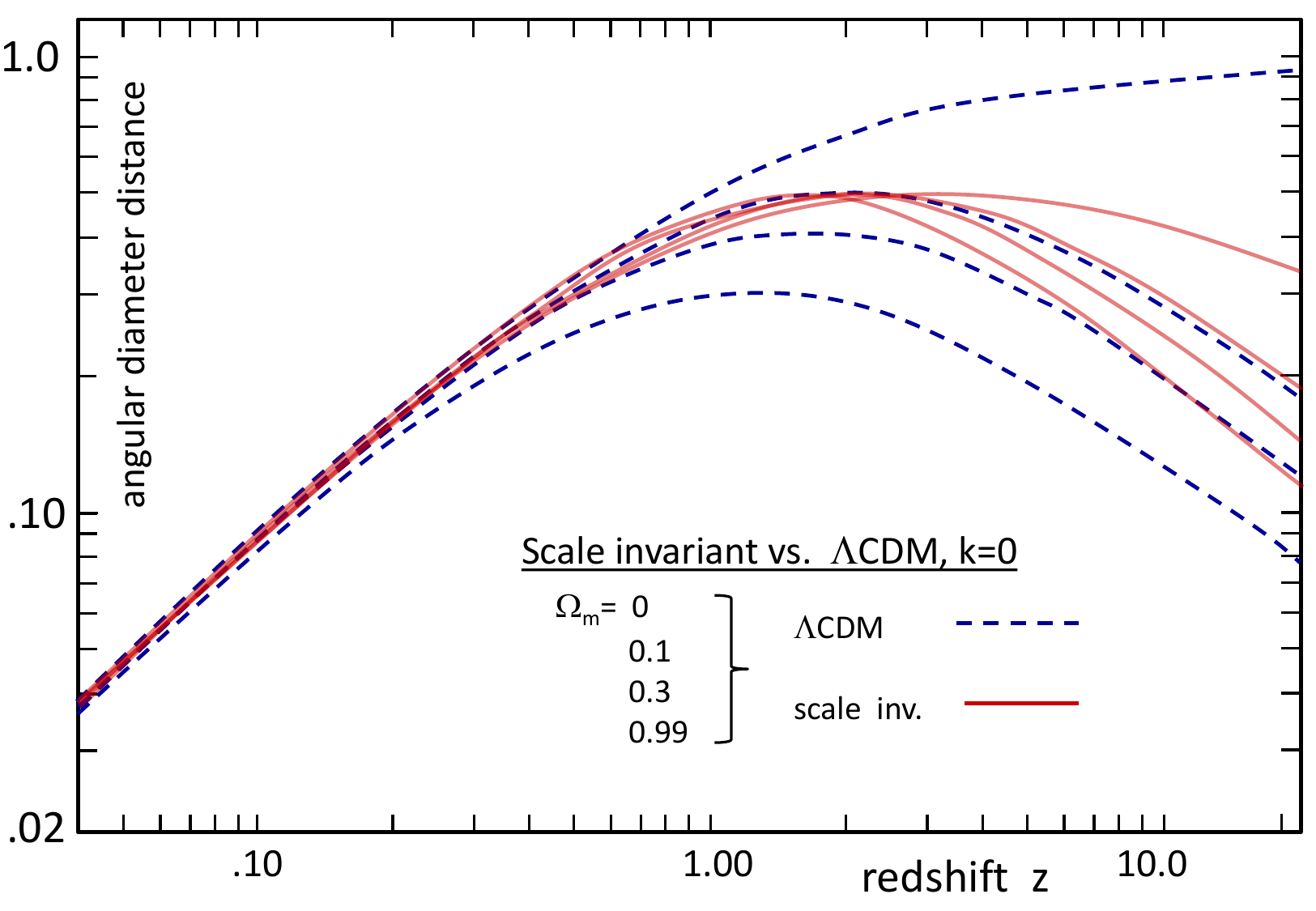}
\caption{\small {\footnotesize{The angular diameter distance $d_{\mathrm{A}}$ vs. redshift $z$ 
for flat scale invariant models  (continuous red lines) compared to flat
$\Lambda$CDM models (broken blue lines).  The  curves are  given for  
$\Omega_{\mathrm{m}} = 0, 0.1, 0.3, 0.99 $,
from the upper to the lower curve  in both cases (at $z  > 3$).} Original figure from \cite{Maeder17a}.}}
\label{distance}
\end{center}
\end{figure}

Most cosmological tests such as  the magnitude-redshift,  the angular diameter vs. redshift, the number count vs. redshifts, etc,
depend on the expressions of the distances based on the angular diameters $d_{\mathrm{A}}$. The plot of 
$d_{\mathrm{A}}$ vs. $z$  in Fig.~\ref{distance} shows that the different curves are  not well separated at lower $z$.
At $z=1$, for $\Omega_{\mathrm{m}} = 0, 0.1, 0.3, 0.99$, one respectively has
$\log d_{\mathrm{A}}= -0.383, -0.367, -0.349, -0.342$. 
Up to a redshift $z=2$, the relations between $d_{\mathrm{A}}$ and $z$ for scale invariant models are very close to each other
whatever  $\Omega_{\mathrm{m}}$, with a deviation from the mean smaller than $\pm 0.05$ dex.
For $\Lambda$CDM models, higher density models always have lower $d_{\mathrm{A}}$
with  an  increasing  separation between the curves with increasing $z$. 
For the SIV models, this is the same, however with a very small differentiation, up to only $z \approx 2$. 
Above $2$, the SIV models behave differently: higher density models have larger  $d_{\mathrm{A}}$ values.
The above properties are evidently also shared by  the magnitude-redshift,  
the angular diameter vs. redshift, as well as by number counts plots.
A clear discrimination  between the SIV and $\Lambda$CDM models with  an access to $\Omega_{\mathrm{m}}$
requires  high precision measurements at redshifts higher than 2.

\begin{figure}[h]
\centering
\includegraphics[width=.85\textwidth]{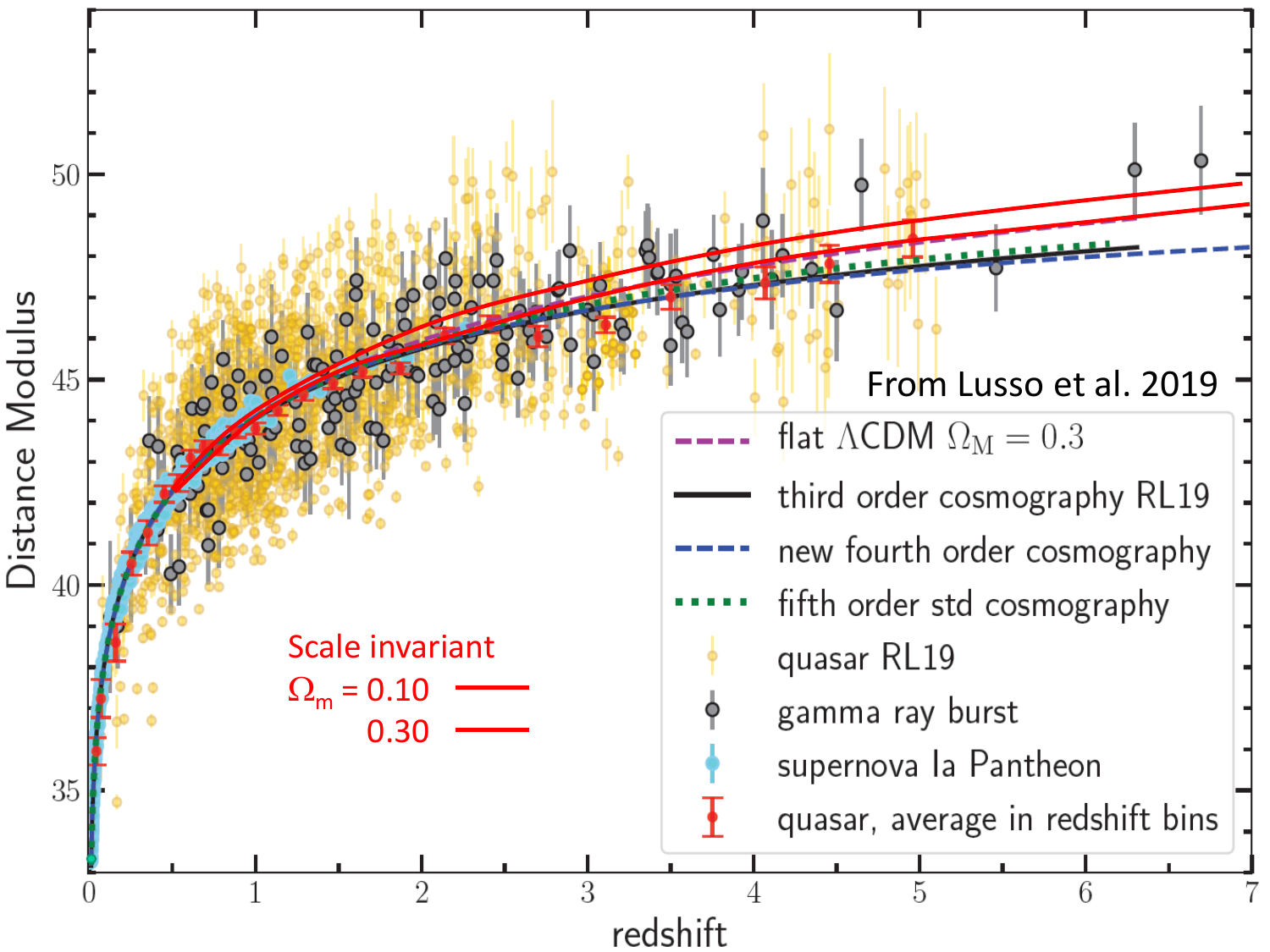}
\caption{\small The Hubble diagram for SNIa, quasars (binned), and GRBs from the samples collected by~\cite{Lusso19}. 
The various models considered by Lusso et al. are indicated. The two red lines show the flat scale invariant models
with $k=0$ and $\Omega_{\mathrm{m}}=0.10$ and 0.30. Note that the $\Lambda$CDM  and SIV models with
$\Omega_{\mathrm{m}}=0.30$ are easily confused. The other lower curves are attempts of adjustments 
by series developments. Drawing originally published in \cite{MaedGueor20a}.}
\label{Lusso}
\end{figure}

Figure~\ref{Lusso} shows  the (m-M) vs. $z$ plot based on  SNIa, quasar, and GRB data  by \cite{Lusso19} compared to 
 different theoretical curves. 
The two red lines show the SIV models for $\Omega_{\mathrm{m}}=0.10$ and 0.30. This last model
lies very close to the $\Lambda$CDM model with $ \Omega_{\mathrm{m}}=0.30$, illustrating the above mentioned difficulty to discriminate
 between the $\Lambda$CDM and SIV models. 
We note that the SIV models  with  $\Omega_{\mathrm{m}}=0.10$  better fits the  high $z$ points, which could perhaps support  a lower value.
However, internal effects in the evolution galaxies may also intervene in the comparison of distant and local galaxies, 
in addition to the cosmological effects and this imposes great care in the conclusions.

Figure~\ref{overall} below, shows the curves of the redshift drifts as a function of  $z$ predicted in the SIV cosmology
for different values of $\Omega_ {\mathrm{m}}$. (A $z$-drift is  the change $z$ for a given galaxy over time,
a time interval longer than 20 yr  appears necessary). The SIV-drifts  are compared to a few standard models of different 
$\Omega_{\Lambda}$-values   by~\cite{Liske08}.
We notice the relative proximity of the standard and scale invariant curves in the case of $\Omega_ {\mathrm{m}}= 0.30$,
which could make the separation of models difficult for such a density parameter. However, 
the expected value of $\Omega_ {\mathrm{m}}$ in the SIV cosmology is likely significantly smaller than in the 
$\Lambda$CDM models; this makes the differences of the $z$-drifts between the 
two kinds of cosmological models possibly observable by very accurate observations in the future.
The physical reason of these differences between the two models at high $z$ is due to the flatter initial expansion curve in the SIV models.
In this respect, we recall that the empty SIV model  expand with $t^2$, while the empty 
$\Lambda$CDM model  is in fact de Sitter model which expands exponentially.

\begin{figure}[h]
\centering
\includegraphics[width=.90\textwidth]{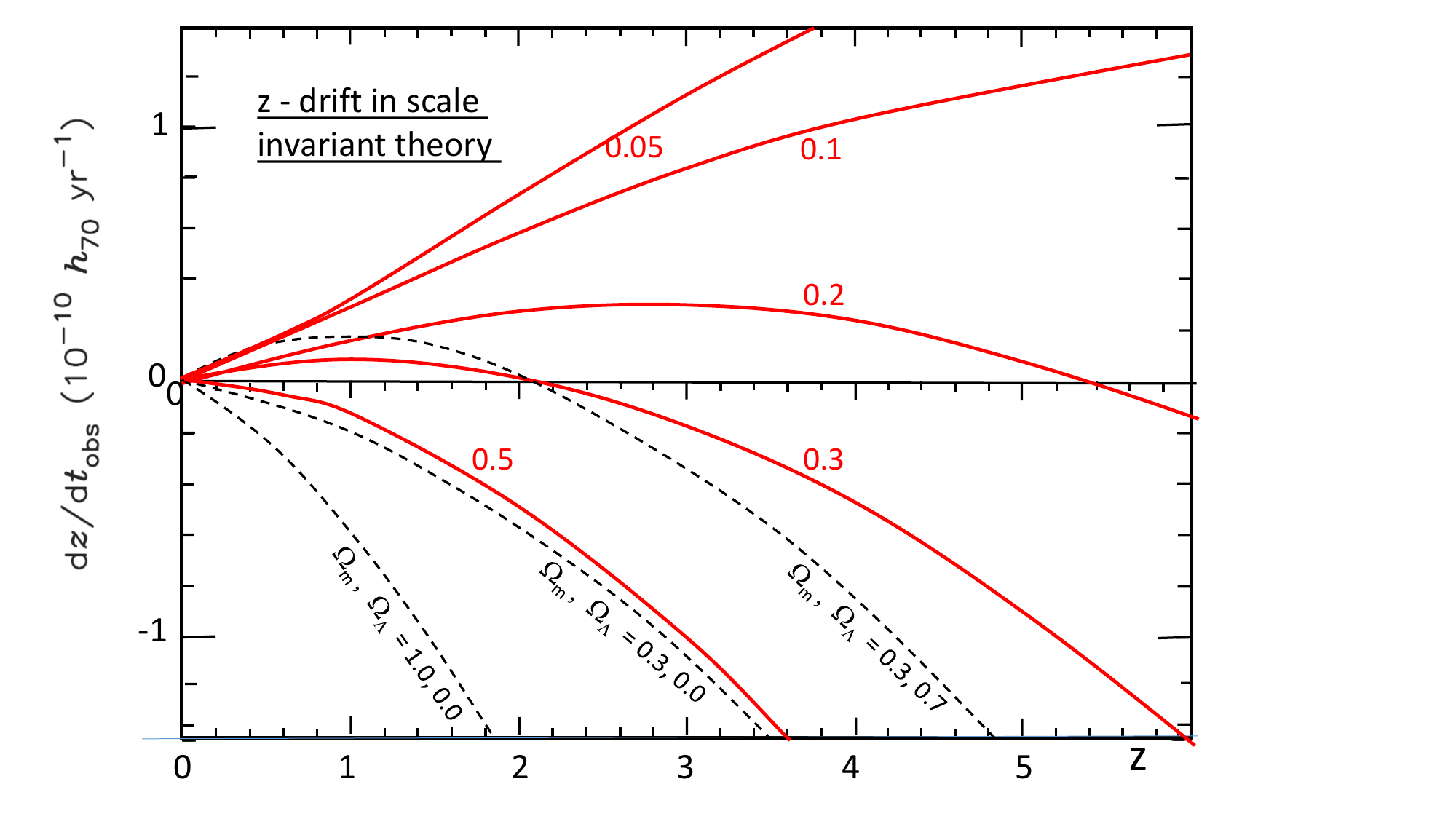}
\caption{\small The drifts of redshifts $dz/dt$ as a function of redshift in the scale invariant theory (red curves). 
The values of $\Omega_{\mathrm{m}}$ 
(usual definition) are indicated. The black broken lines give the results for some standard models of different couples 
$(\Omega_{\Lambda}, \Omega_{\mathrm{m}})$ by~\cite{Liske08}.
Drawing originally published in \cite{MaedGueor20a}.}
\label{overall}
\end{figure} 

The above comparisons, see also \cite{MaedGueor20a}, 
show a general agreement between SIV predictions and observations, alike for the $\Lambda$CDM models.
The  redshift drifts appear to have a particularly great   differentiation power between SIV and $\Lambda$CDM models.

\subsection{Scale-Invariant Dynamics of Galaxies \cite{MaedGueor20b}}

The next important application of the scale-invariance at cosmic scales is
the derivation of a universal expression for the
Radial Acceleration Relation (RAR) of $g_{\mathrm{obs}}$  and $g_{\mathrm{bar}}$.
That is, the relation between the observed gravitational acceleration $g_{\mathrm{obs}}=v^2/r$
and the acceleration from the baryonic matter due to the 
standard Newtonian gravity $g_\mathrm{N}$ \cite{MaedGueor20b} ($g=g_{\mathrm{obs}}$, $g_N=g_{\mathrm{bar}}$):
\begin{equation}
g\,=\,g_{\mathrm{N}}+\frac{k^{2}}{2}+\frac{1}{2}\sqrt{4g_{\mathrm{N}}k^{2}+k^{4}}\,,
\label{sol}
\end{equation}
For $g_{\mathrm{N}} \gg k^{2} : g \rightarrow g_{\mathrm{N}}$
but for $g_{\mathrm{N}}\rightarrow 0 \Rightarrow g \rightarrow k^{2}$ is a constant.
\begin{figure}[h]
\centering 
\includegraphics[width=0.75\textwidth]{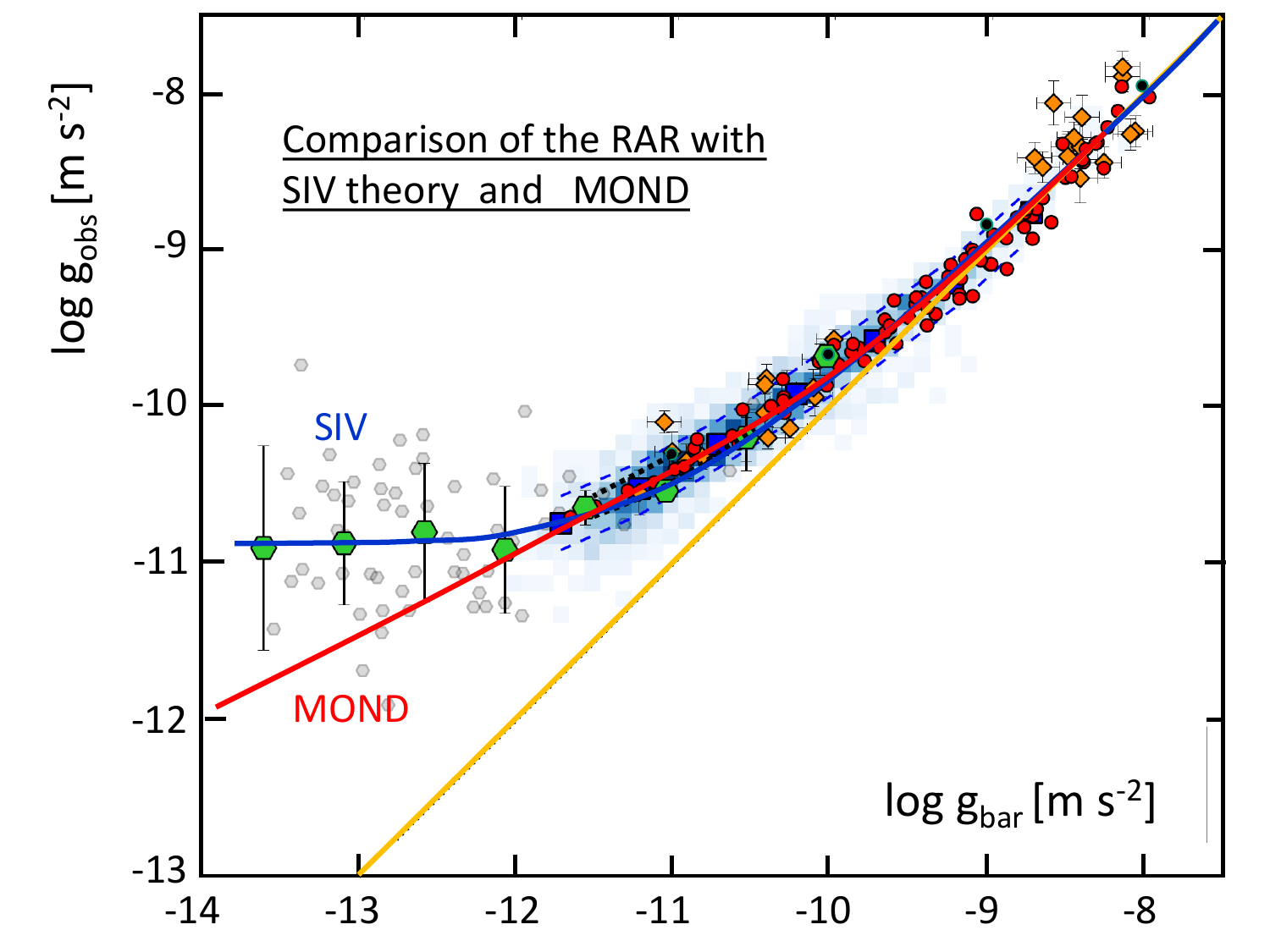} 
\caption{\small 
Radial Acceleration Relation (RAR) for the galaxies studied by Lelli et al. (2017).
Dwarf Spheroidals  as binned data (big green hexagons), 
along with MOND (red curve), and SIV (blue curve) model predictions.
The orange curve shows the 1:1-line for $g_{\mathrm{obs}}$  and $g_{\mathrm{bar}}$.
Due to the smallness of  $g_{\mathrm{obs}}$  and $g_{\mathrm{bar}}$ 
the application of the $\log$ function results in negative numbers; 
thus, the corresponding axes' values are all negative.
Drawing originally published in \cite{MaedGueor20b}.}
\label{gobs} 
\end{figure}

As shown in Fig.~\ref{gobs}, MOND significantly deviates from the trend seen in the Dwarf Spheroidals. 
This is a problem within MOND due to the different interpolating functions needed by MOND -
one for galaxies and one for cosmic scales. 
The SIV expression \eqref{sol}  naturally resolves this issue  by using only one universal parameter  
related to gravity at large distances \cite{MaedGueor20b}.

The above equation \eqref{sol} is derived from the Weak Field Approximation of the SIV by 
using the Dirac co-calculus in the derivation of the geodesic equation within the relevant WIG \eqref{eq:geodesics*}. 
For more details, see \cite{MaedGueor20b} and the original derivation in \cite{MBouvier79}:

\begin{eqnarray}
g_{ii}=-1,\;g_{00}&=&1+2\Phi/c^{2}\Rightarrow\Gamma_{00}^{i}
=\frac{1}{2}\frac{\partial g_{00}}{\partial x^{i}}=\frac{1}{c^{2}}\frac{\partial\Phi}{\partial x^{i}},\nonumber\\
\frac{d^{2}\overrightarrow{r}}{dt^{2}}&=&-\frac{G_tM}{r^{2}}\frac{\overrightarrow{r}}{r}
+\kappa_{0}(t)\frac{d\overrightarrow{r}}{dt}.\label{Nvec}
\end{eqnarray}
where $i\in{1,2,3}$, while the potential $\Phi=G_tM/r$ is scale invariant and 
$G_t$ is the Newton's constant of gravity in SIV $t$-time units system ($t_0=1$) . 
When written in the usual units with present time $\tau_0$,  
based on \eqref{kappa00} the modified Newton's equation \eqref{Nvec} is then \cite{Maeder23,2022arXiv220413560M}:
\begin{equation}
\frac {d^2 \overrightarrow{r}}{d \tau^2}  \, = 
\, - \frac{G \, M(\tau_0)}{r^2} \, \frac{\overrightarrow{r}}{r}   + \frac{\psi_0}{\tau_0}   \frac{d\overrightarrow{r}}{d\tau} \, \,.  
\label{Nvec4}
\end{equation}

By considering the scale-invariant ratio of the correction term $\kappa_0 \, \upsilon\, $ 
to the usual Newtonian term in \eqref{Nvec}, one has:
\begin{equation}
x=\frac{\kappa_0 \upsilon r^{2}}{GM}=\frac{H_{0}}{\xi}\frac{\upsilon \,r^{2}}{GM}
= \, \frac{H_0}{\xi}  \frac{(r \, g_{\mathrm{obs}})^{1/2}}{g_{\mathrm{bar}}}
\sim\frac{g_{\mathrm{obs}}-g_{\mathrm{bar}}}{g_{\mathrm{bar}}}\,,
\label{x}
\end{equation}

\noindent
Upon utilizing an explicit scale invariance, by considering ratios, for canceling the proportionality factor, we obtain:
\begin{equation}
{\left(\frac{g_{\mathrm{obs}}-g_{\mathrm{bar}}}{g_{\mathrm{bar}}}\right)_{2}}\div
{\left(\frac{g_{\mathrm{obs}}-g_{\mathrm{bar}}}{g_{\mathrm{bar}}}\right)_{1}}\,=
\,\left(\frac{g_{\mathrm{obs, 2}}}{ g_{\mathrm{obs, 1}}}\right)^{1/2}\,
\left(\frac{g_{\mathrm{bar,1}}}{g_{\mathrm{bar,2}}}\right)\,,
\label{corrg}
\end{equation}
by setting $g=g_{\mathrm{obs, 2}}$, $g_N=g_{\mathrm{bar,2}}$, 
and with $k=k_{(1)}$ containing all the system-1 terms, one finally obtains \eqref{sol}:
\begin{equation*}
\frac{g}{g_{\mathrm{N}}}-1=k_{(1)}\frac{g^{1/2}}{g_{\mathrm{N}}}
\Rightarrow
g\,=\,g_{\mathrm{N}}+\frac{k^{2}}{2}\pm\frac{1}{2}\sqrt{4g_{\mathrm{N}}k^{2}+k^{4}}.
\end{equation*}
As it was noticed already,  $g_{\mathrm{N}} \gg k^{2} : g \rightarrow g_{\mathrm{N}}$
but for $g_{\mathrm{N}}\rightarrow 0 \Rightarrow g \rightarrow k^{2}$ for the `$+$' branch, 
while the `$-$' branch gives $g \rightarrow 0$.
\subsection{MOND as a peculiar case of the SIV theory \cite{Maeder23}}
The weak field limit of SIV  tends to MOND, when the scale factor is taken as constant, 
an approximation valid  ($<1\%$) over the last 400 Myr.
A better understanding  of the MOND $a_0$-parameter in $g_\text{obs}=\sqrt{a_0\,g_N}$ could be obtained 
within the SIV where it corresponds to the equilibrium point of the Newtonian and  SIV dynamical  acceleration \cite{Maeder23}; 
as such, the parameter $a_0$ is not a universal constant, it depends on the density and  age of the Universe.

In order to see the correspondence one looks at $x\xi=H_{0}\frac{\upsilon \,r^{2}}{GM}$ \eqref{x}  in terms of densities:
first consider the mass $M$  spherically distributed in a radius $r$ with a mean density $\varrho=3M/(4\pi\,r^3)$,
then use  $\varrho_{\mathrm{c}}= \frac{3 \, H_0^2}{8 \, \pi \,G}\Rightarrow\,H_0=\sqrt{8\pi\,G\rho_c/3}$, 
along with the instantaneous radial accelerator relation 
$\frac{\upsilon^2}{r} = \frac{GM}{r^2}\Rightarrow\,\upsilon=\sqrt{GM/r}$,
to arrive at the expression  $x\xi=\sqrt{2\rho_c/\rho}$.
Since Newtonian gravity  for a density $\rho$ is $g_N= (4/3) \pi G \, \varrho \, r$ this translates into 
$x\xi=\sqrt{2g_c/g_N}$. Then one can write (\ref{Nvec4}) as
 \begin{equation}
 g \, =  \, g_{\mathrm{N}} + x \, g_{\mathrm{N}} \,\rightarrow x \, g_{\mathrm{N}}=
\frac{\sqrt{2}}{\xi} \,\left(\frac{g_{\mathrm{c}}}{g_{\mathrm{N}}}\right)^{1/2} g_{\mathrm{N}}\,=
\frac{1}{\xi} \,\sqrt{2g_{\mathrm{c}} g_{\mathrm{N}}}.
 \end{equation}
 \noindent
Therefore, one has the correspondence
$a_0 \, \Longleftrightarrow  \, 2 g_{\mathrm{c}}/{\xi^2},$ where $\xi=H_0/\kappa_0$.
Thus, by using the SIV-time  scale $t$, where $\kappa_0(t)=1/t$ due to  \eqref{lambda(t)}, 
along with \eqref{HM}, one has $\xi=2/(1-\Omega_{\mathrm{m}})$ which finally gives:
\begin{equation}
a_0 \, \Longleftrightarrow  \, \frac{(1-\Omega_{\mathrm{m}})^2}{2}  g_{\mathrm{c}} \, .
\label{aog1}
\end{equation}

One may express the limiting value  $  g_{\mathrm{c}}$ in term of the critical density 
over the radius $r_{\mathrm{H_0}}$ of the Hubble sphere.
Thus, $r_{\mathrm{H_0}}$ is defined via $ n\, c \, = r_{\mathrm{H_0}} H_0$ where $n$ depends on the cosmological model. 
For the EdS model $n=2$, while for SIV or $\Lambda$CDM  models with $\Omega_ {\mathrm{m}}= 0.2 - 0.3$, 
the initial braking and recent acceleration almost compensate each other, so that  $n \simeq 1$. 
By using the expression for $H_0$ \eqref{H0} one finally  obtains:
 \begin{equation} 
 a_0 \, = \, \frac{(1-\Omega_{\mathrm{m}})^2}{2} \, \frac{4  \pi}{3} G  \varrho_{\mathrm{c}} r_{\mathrm{H_0 }} =
 \frac{(1-\Omega_{\mathrm{m}})^2}{4} \, n\,  c \, H_0\,
= \, \frac{n\, c \, (1-\Omega_{\mathrm{m}}) (1-\Omega^{1/3}_{\mathrm{m}})}{2\, \tau_0}.
 \label {aog}
 \end{equation}

Thus, the deep-MOND limit is found \cite{Maeder23} to be  an  approximation of the SIV theory for low enough densities 
and for systems with timescales  smaller than a few Myr where $\lambda$ can be viewed as if it is a constant.  

 The product $c \, H_0$ is equal to 6.80 $\cdot10^{-8}$ cm s$^{-2}$.
 For $\Omega_{\mathrm{m}}$=0, 0.10, 0.20, 0.30 and 0.50, 
 one has $a_0 \approx$ (1.70, 1.36, 1.09, 0.83, 0.43) $\; \cdot \;  10^{-8}$ cm s$^{-2}$ respectively.
 These values obtained from the SIV theory are remarkably close to the value 
   $a_0$ about 1.2 $\cdot \,10^{-8}$ cm s$^{-2}$  derived from  observations by \cite{Milgrom15}. 
 
 Thus, as it comes out for the more general SIV theory, 
 there are several remarks to be made on the $a_0$-parameter and its meaning:
 
\begin{enumerate}

\item The  equation of the deep-MOND limit is reproduced by the SIV theory both  analytically  and numerically 
if $\lambda$ and $M$ can be considered as  constant. This may apply to systems with a typical dynamical 
timescale  up to a few hundred million years.
  
\item  Parameter $a_0$ is not a universal constant.  It depends  on the Hubble-Lema\^{i}tre $H_0$ parameter 
(or  the age  of the Universe) and on $\Omega_{\mathrm{m}}$ in the model  Universe, 
cf. Eq. (\ref{aog}). The value of $a_0$  applies to the present epoch.  

\item Parameter $a_0$ is defined by the condition that $x >1$, {\emph{i.e.}} when the dynamical 
gravity  $\kappa_0\upsilon=(\psi_0 \upsilon)/ \tau_0$  in the equation of motion  (\ref{Nvec4}) becomes larger than the Newtonian gravity.
This situation occurs in regions at the edge of gravitational systems.

\end{enumerate}

\subsection{Local dynamical effects within SIV - the lunar recession \cite{2022arXiv220413560M}}

We have already pointed out that scale invariance  is  expected in empty Universe models, 
while the presence of matter tends to suppress it.
Scale invariance is certainly absent in cosmological models with densities 
equal to or above the critical value $\varrho_{\mathrm{c}} =3H^2_0/(8 \pi G)$ \cite{SIV-Inflation'21}.
Clearly,  the presence of matter tends to kill  scale invariance as shown by \cite{Feynman63}.
For models with densities below $\varrho_{\mathrm{c}}$, 
the possibility of limited  effects remains open.
If present, scale invariance would  be  a global cosmological
property.  Some traces could be observable locally. For the Earth-Moon two-body system, 
the predicted additional lunar recession would be increased by 0.92 cm/yr, 
while the tidal interaction would  also be slightly increased  \cite{2022arXiv220413560M}.

The Earth-Moon distance is the most systematically  measured distance in the Solar System, thanks to
the Lunar Laser Ranging  (LLR) experiment   active since 1970. 
The observed lunar recession  from LLR  amounts to 3.83 ($\pm 0.009$) cm/yr;
implying a  tidal change   of the length-of-the-day (LOD) by 2.395 ms/cy  \cite{Williams16a,Williams16b}.
The value of the lunar recession has not much changed since the first determination  
more than  three decades ago \cite{Christodoulidis88}, which illustrates the  quality of the measurements.
However,  the observed change of the LOD  
since the Babylonian Antiquity is only 1.78 ms/cy  \cite{Stephenson16}, 
a result supported by paleontological data
Deines and Williams \cite{Deines16}, 
and implying  a lunar recession of  2.85 cm/yr.
The best and longest studies on the change of the LOD in History 
have been performed by Stephenson et al.  \cite{Stephenson16},
who  analyzed the lunar and solar eclipses from 720 BC up to 1600 AD 
and found an average  shift of the LOD by 1.78 ($\pm$ 0.03) ms/cy.
The reality of the difference between the above observed mean value of the LOD (1.78 ms/cy) 
and the value due to the tidal interaction (2.395 ms/cy) has been further emphasized by \cite{Stephenson20}.  

The significant difference  of  (3.83-2.85) cm/yr = 0.98 cm/yr,
already pointed out by several authors  over the last two decades \cite{McCarthy86, Sidorenkov97}, 
corresponds well to the predictions of the scale-invariant theory, 
which is also supported by  several other astrophysical  tests \cite{2022arXiv220413560M}. 

By using the correct treatment of the Earth-Moon tidal interaction within the  SIV theory one 
derives an additional terms in the equation describing the lunar recession  
in current time units \cite{2022arXiv220413560M}: 
\begin{equation}
\frac{dR}{d\tau} \, = \, k_{\mathrm{E}} \, \frac{dT_{\mathrm{E}}}{d\tau}- 
 k _{\mathrm{E}} \,   \psi _0\,\frac{T_{\mathrm{E}}}{\tau_0} +
 \psi_0  \,\frac{R}{\tau_0}\, .
\label{fin}
\end{equation}
In a cosmological model with $\Omega_{\mathrm{m}}=0.30$, 
the ratio $\psi_0 = \frac{(t_0-t_{\mathrm{in}})}{t_0} \,=0.331$ (\ref{kappa00}).
We use the following numerical values of the relevant astronomical quantities:
\begin{eqnarray}
M_{\mathrm{E}} = 5.973 \cdot 10^{27} \mathrm{g},  \quad  \quad R_{\mathrm{E}}= 6.371 \cdot 10^{8} \mathrm{cm}, \\ \nonumber
\; \; \quad M_{\mathrm{M}}= 7.342 \cdot 10^{25} \mathrm {g},\quad \quad R=  3.844 \cdot 10^{10} \mathrm{cm},\\ \nonumber
\quad I_{\mathrm{E}} = 0.331 \cdot  M_{\mathrm{E}}   R^2_{\mathrm{E}} =
8.0184 \cdot  10^{44} \mathrm{g \cdot cm}^2. \nonumber
\end{eqnarray}
The value 0.331 is obtained from precession data \cite{Williams94}.
The coefficient $k_{\mathrm{E}}$ is estimated to be $1.60 \cdot 10^5$ cm $\cdot$ s$^{-1 }$ 
\cite{2022arXiv220413560M, 2021Ap&SS.366..101M}. 

Let us evaluate numerically the various   contributions. 
With the LOD  of 1.78 ms/cy from the antique data by \cite{Stephenson16}, the first term contributes to a lunar recession of 2.85 cm/yr.
The second term  in (\ref{fin}) gives for the case of $\Omega_{\mathrm{m}}=0.3$,
\begin{equation}
0.33 \cdot k_E \frac{T_{\mathrm{E}}}{\tau_0} \, = \, 0.33 \cdot 1.60 \cdot 10^5 cm \cdot s^{-1 } \; \frac{86400 \; s}{13.8 \cdot 10^9 \; yr}
 = 0.33 \; \left[\rm \frac{cm}{yr}\right] \, .
 \end{equation}
The direct SIV expansion effect $\kappa_0R=\psi _0 R/\tau_0$ is
\begin{equation}
0.33 \cdot  \frac{R}{\tau_0} \, = \,0.33 \cdot  \frac{3.844 \cdot 10^{10} \, cm}{13.8 \cdot 10^9 \; yr} = 0.92  \; 
\left[\rm \frac{cm}{yr}\right] \, .
\end{equation}
This term corresponds to a third of the general Hubble-Lemaitre expansion.
Summing the various contributions, we get the historical data value \cite{Stephenson16}:
\begin{eqnarray}
\frac{dR}{d\tau}  = (2.85-0.33+0.92) \,\mathrm{ cm/yr}
= 3.44 \,\mathrm{cm/yr}.
\label{final}
\end{eqnarray}
Thus, we see that the scale invariant analysis is giving a relatively good agreement with the  lunar recession
 of 3.83 cm/yr obtained  from LLR observations. The difference amounts only to  10 \% of the observed lunar recession. 

 The difference in the lunar recession is well accounted for within the dynamics of the  SIV theory (\ref{final}).  
 \emph{A minima},  the above results shows that the problem of scale invariance is worth of some attention
 within the solar system as well.

\subsection{Growth of the Density Fluctuations within SIV \cite{MaedGueor19}}

\begin{figure}[h]
\centering 
\includegraphics[width=0.6\textwidth]{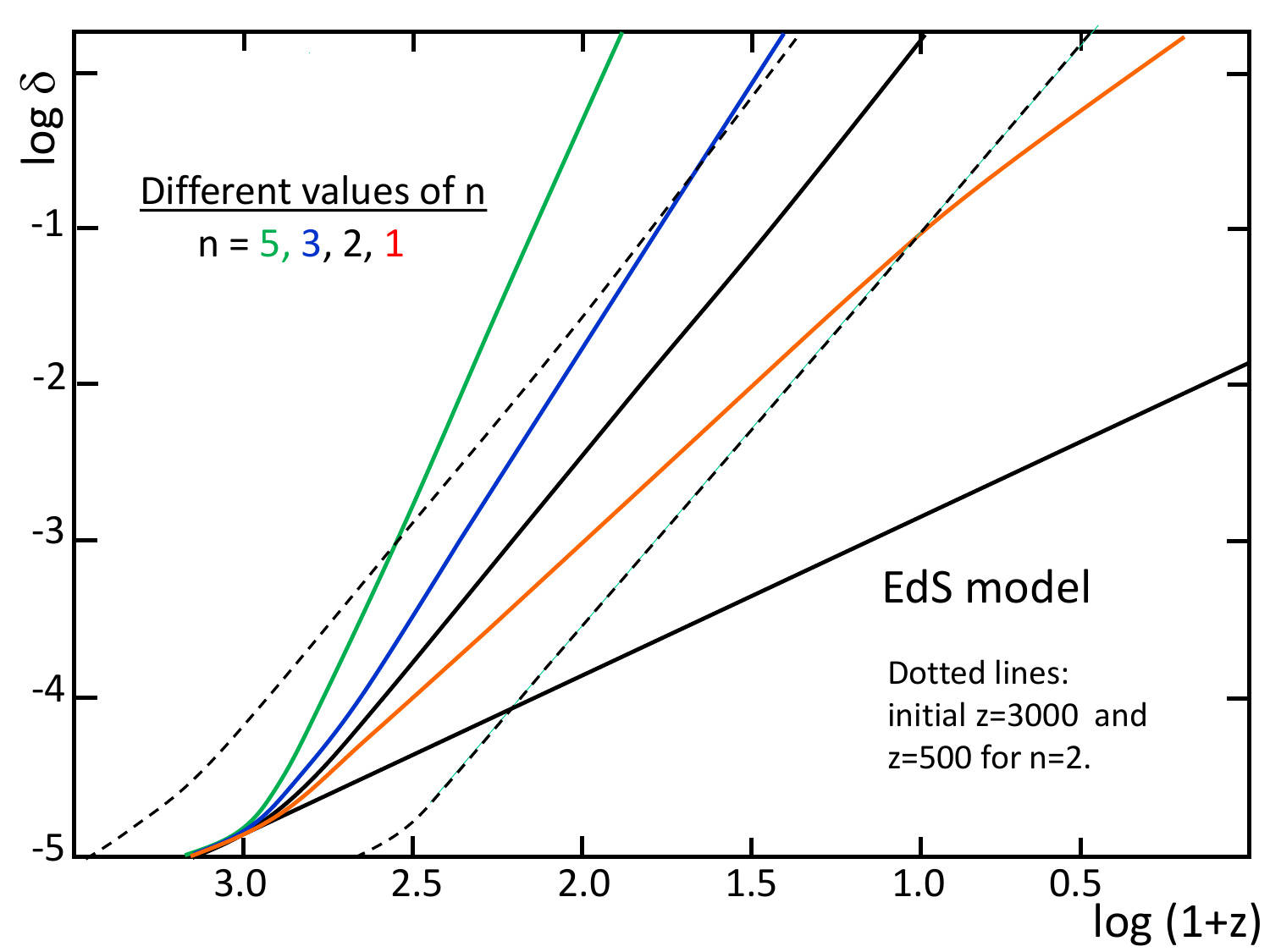} 
\caption{\small 
The growth of density fluctuations for different values of parameter
$n$ (the gradient of the density distribution in the nascent cluster),
for an initial value $\delta=10^{-5}$ at $z=1376$ and $\Omega_{\mathrm{m}}=0.10$.
The initial slopes are those of the EdS models. The two light broken
curves show models with initial $(z+1)=3000$ and 500, with same $\Omega_{\mathrm{m}}=0.10$
and $n=2$. These dashed lines are to be compared to the black continuous
line of the $n=2$ model. All the three lines for $n=2$ are very
similar and \mbox{nearly parallel}.
Due to to the smallness of  $\delta$  the application of the $\log$ function results in negative numbers; 
thus, the corresponding vertical axes values are all negative.
Drawing originally published in \cite{MaedGueor19}.}
\label{variousn} 
\end{figure}

Another interesting finding was the potential for rapid growth of density fluctuations within the SIV \cite{MaedGueor19}. 
Our study appropriately modified the relevant equations: 
the continuity equation, Poisson equation, and Euler equation, within the SIV framework. 
Below are outlined the corresponding equations and the relevant results. 
Using the notation $\kappa=\kappa_0=-\dot{\lambda}/\lambda=1/t$, 
the corresponding Continuity, Poisson, and Euler equations are:

\begin{eqnarray*}
\frac{\partial \rho}{\partial t}+ \vec{\nabla}\cdot (\rho \vec{v}) 
= \kappa \left [\rho+ \vec{r} \cdot \vec{\nabla} \rho \right]\,,\; 
\vec{\nabla}^{2}\Phi=\triangle\Phi=4\pi G \varrho \label{eq:Continuity+Poisson},\\
\frac{d\vec{v}}{dt}=\frac{\partial\vec{v}}{\partial t}+\left(\vec{v}\cdot\vec{\nabla}\right)\vec{v}=
-\vec{\nabla}\Phi-\frac{1}{\rho}\vec{\nabla}p+\kappa\vec{v}\label{Euler} \, .
\end{eqnarray*}

\noindent
For a density perturbation $\varrho(\vec{x},t)\, = \, \varrho_{b}(t)(1+\delta(\vec{x},t))$
the above equations result in:
\begin{eqnarray}
\dot{\delta} + \vec{\nabla} \cdot \dot{\vec{x}} =\kappa \vec{x} \cdot \vec{\nabla} \delta = n \kappa(t) \delta &,&
\vec{\nabla}^{2}\Psi = 4\pi Ga^{2}\varrho_{b}\delta,\label{D1}\\
\ddot{\vec{x}}+ 2 H \dot{\vec{x}}+ (\dot{\vec{x}}\cdot \vec{\nabla}) \dot{\vec{x}} 
&=& -\frac{\vec{\nabla} \Psi}{a^2} +\kappa(t) \dot{\vec{x}}.\\
\Rightarrow \ddot{\delta}+ (2H -(1+n) \kappa )\dot{\delta} 
&=& 4\pi G \varrho_{b}\delta + 2 n \kappa (H-\kappa) \delta.\;\;
\label{D2}
\end{eqnarray}

The above equation \eqref{D2}, which is the final result of \cite{MaedGueor19}, 
reduces to the standard equation when $\kappa$ approaches 0. 
The simplifying assumption in equation \eqref{D1}, 
introduces the parameter $n$ that measures the type of the perturbation.
\footnote{ The perturbation type number $n$ in our study is related 
to the degree of the homogeneous polynomial that describes the constant hyper-surfaces of  $\delta$.
That is, a linear perturbations are represented with $\delta=x^i\,a_i$ for constants $a_i$, 
while   $\delta=x^i\,x^j\,a_{ij}$ will correspond to quadratic polinomials and so on.} 

That is, a spherically symmetric perturbation has $n=2$. 
As shown in Fig.~\ref{variousn}, even at relatively low matter densities, 
perturbations for various values of $n\ge1$ result in faster growth of 
the density fluctuations than in the Einstein--de Sitter model. 
Strikingly, the overall slope is independent of the choice of recombination epoch $z_\mathrm{rec}$.

\begin{figure}[h]
\centering 
\includegraphics[width=0.6\textwidth]{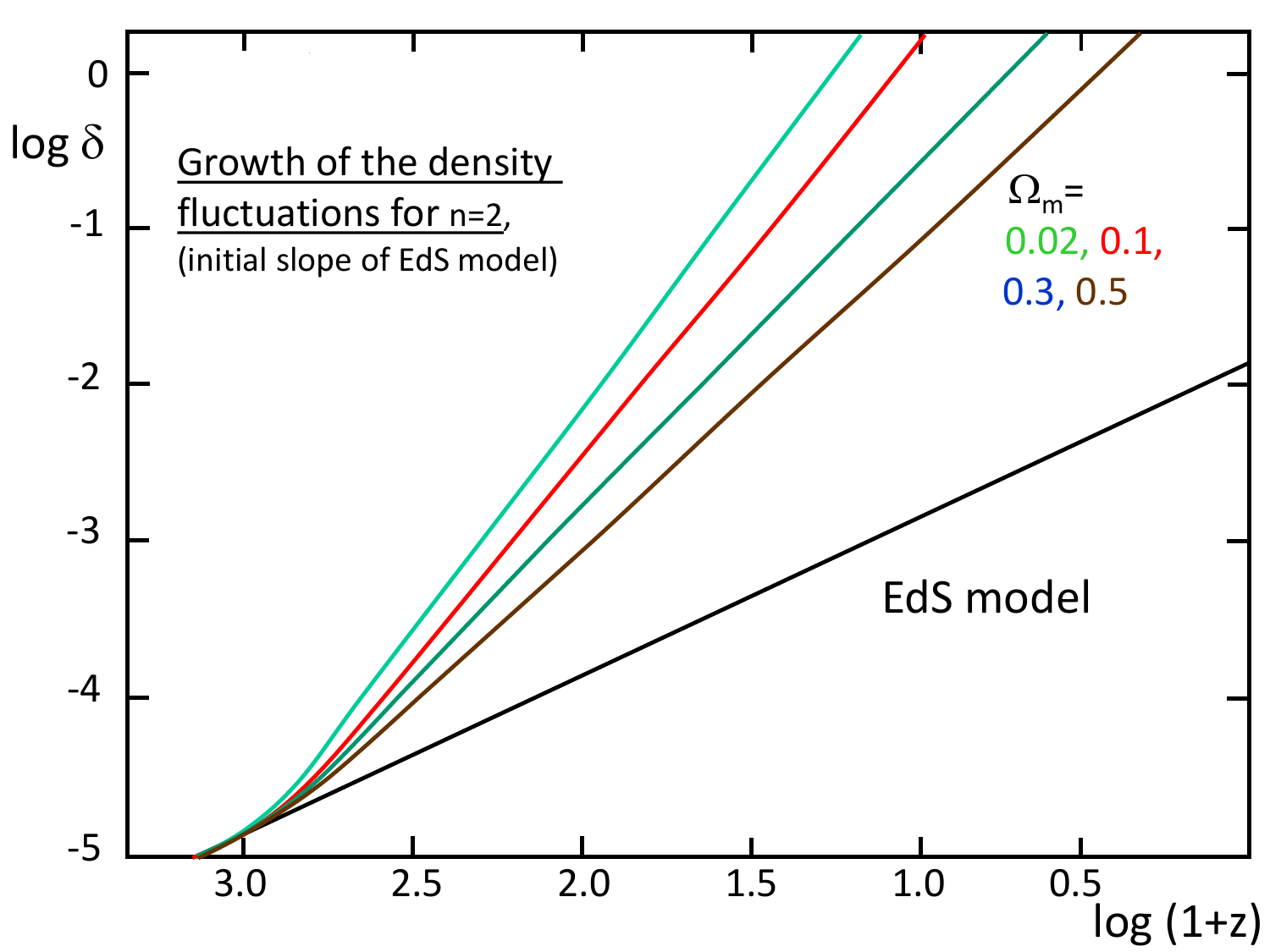} 
\caption{\small The black curve is the classical growth of the density fluctuations in the Einstein-de Sitter model. 
The other four curves illustrate the growth of $\delta$ for the density profile with $n=2$ in the scale-invariant theory. 
There are four different values of the density parameter $\Omega_{\mathrm{m}}$. 
An initial value $\delta= 10^{-5}$ at $z=1376$ has been taken for all models, the initial derivative 
$\dot{\delta}$ is taken equal to that of the EdS model. After a short evolution with a slope close 
to the initial one, all solutions indicate a much faster growth of the density fluctuations,
reaching the non-linear regime between about $z+1= 2.7$ and $z=18$.
Drawing originally published in \cite{MaedGueor19}.}
\label{EdSn2} 
\end{figure}

The behavior for different $\Omega_\mathrm{m}$ is also interesting (see Fig. \ref{EdSn2}), 
for example, the smaller  $\Omega_\mathrm{m}$ is - the steeper the 
growth of the density fluctuations is. It is always much steeper then the Einstein-de~Sitter model.
For further details see the discussion by \cite{MaedGueor19}.

Over the recent years highly redshifted galaxies have been found, 
in particular with the observational data from JWST,  
which is suggesting very early times of galaxy formation \cite{Bunker'23}.  We point out, as shown by Fig. \ref{variousn} , 
that very early galaxy formation is a process currently expected in the context of the SIV theory.

\subsection{Big-Bang Nucleosynthesis within SIV  \cite{VG&AM'23}}

The SIV paradigm has been recently applied to the Big-Bang Nucleosynthesis 
using the known analytic expressions for the expansion factor $a(t)$ and
the plasma temperature $T$ as functions of the SIV time $\tau$ since the Big-Bang
when $a(\tau=0)=0$ \cite{VG&AM'23}. The results have been compared to the standard
BBNS as calculated via the PRIMAT code \cite{PRIMAT}. Potential SIV-guided deviations from the 
local statistical equilibrium were also explored in \cite{VG&AM'23}. 
Overall, it was found that smaller than usual baryon 
and non-zero dark matter content, by a factor of three to five times reduction, 
result in compatible to the standard light elements abundances (Table \ref{Table2}).

\begin{table}[h]
\centering
\scalebox{0.8}{
\begin{tabular}{|c|cc|ccc|ccc|}
\hline\hline 
Element & Obs. & PRMT & $a_{SIV}$ & fit & fit* & $\bar{a}/\lambda$&  fit* & fit \\
\hline 
 \text{H} & 0.755 & 0.753 & 0.805 & 0.755 & 0.849 & 0.75 & 0.753 & 0.755 \\
 $Y_P=4Y_{\text{He}}$ & 0.245 & 0.247 & 0.195 & 0.245 & 0.151 & 0.25 & 0.247 & 0.245 \\
 $\text{D/H}\times10^5$ & 2.53 & 2.43 & 0.743 & 2.52 & 2.52 & 1.49 & 2.52 & 2.53 \\
\hline  
 $^3\text{He/H}\times10^5$ & 1.1 & 1.04 & 0.745 & 1.05 & 0.825 & 0.884 & 1.05 & 1.04 \\
 $^7\text{Li/H}\times10^{10}$ & 1.58 & 5.56 & 11.9 & 5.24 & 6.97 & 9.65 & 5.31 & 5.42 \\
\hline  
$N_{\text{eff}}$ & 3.01 & 3.01 & 3.01 & 3.01 & 3.01 & 3.01 & 3.01 & 3.01 \\
$\eta_{10}$ & 6.09 & 6.14 & 6.14 & 1.99 & 0.77 & 1.99 & 5.57 & 5.56 \\
 FRF  & 1 & 1 & 1 & 1 & 1.63 & 1 & 1 & 1.02 \\
 \text{m\v{T}} & 1 & 1 & 1 & 1 & 0.78 & 1 & 1 & 0.99 \\
 \text{Q/\v{T}} & 1 & 1 & 1 & 1 & 1.28 & 1 & 1 & 1.01 \\
$\Omega _b$\; [\%] & 4.9 & 4.9 & 4.9 & 1.6 & 0.6 & 1.6 & 4.4 & 4.4 \\
 $\Omega _\mathrm{m}$\;[\%]  & 31 & 31 & 31 & 5.9 & 23 & 5.9 & 86 & 95 \\
 $\sqrt{\chi _{\epsilon}^2}$ & N/A & 6.84 & 34.9 & 6.11 & 14.8 & 21.9 & 6.2 & 6.4 \\
\hline\hline 
\end{tabular}}
\caption{\small \label{Table2}
The observational uncertainties are  1.6\% for $Y_{P}$, 1.2\% for D/H, 18\% for T/H, and 19\% for Li/H.
FRF is the forwards rescale factor for all reactions, 
while \text{m\v{T}}  and \text{Q/\v{T}} are the corresponding rescale factors
in the revers reaction formula based on the local thermodynamical equilibrium. 
The SIV $\lambda$-dependences are used when these factors are different from 1;
that is, in the sixth and ninth columns where FRF=$\lambda$, 
\text{m\v{T}}= $\lambda^{-1/2}$, and \text{Q/\v{T}}= $\lambda^{+1/2}$. 
The columns denoted by fit contain the results for perfect fit 
on $\Omega _b$ and $\Omega _\mathrm{m}$  to $^{4}$He and D/H,
while fit* is the best possible fit on $\Omega _b$ and $\Omega _\mathrm{m}$ 
to the $^{4}$He and D/H observations 
for the model considered as indicated in the columns four and seven.
The last three columns are usual PRIMAT runs with modified $a(T)$ 
such that $\bar{a}/\lambda=a_{SIV}/{S}^{1/3}$,
where $\bar{a}$ is the PRIMAT's $a(T)$ for the decoupled neutrinos case.
Column seven is actually $a_{SIV}/{S}^{1/3}$,
but it is denoted by $\bar{a}/\lambda$ to remind us 
about the relationship $a'=a\lambda$;
the run is based on $\Omega _b$ and $\Omega _\mathrm{m}$ from column five.
The smaller values of $\eta_{10}$ are due to smaller 
$h^2\Omega_b$, as seen by noticing that 
$\eta_{10}/\Omega_b$ is always $\approx 1.25$.
Table originally presented in \cite{VG&AM'23}.}
\end{table}

The SIV analytic expressions for $a(T)$ and $\tau(T)$ were utilized to study the BBNS within the SIV paradigm \cite{VG&AM'23, Maeder19}.
The functional behavior is very similar to the standard model within PRIMAT except during the very early Universe 
where electron-positron annihilation and neutrino processes affect the $a(T)$ function (see Table I and Fig. 2 in \cite{VG&AM'23}). 
The distortion due to these effects encoded in the function $S(T)$ could be incorporated by considering the SIV paradigm 
as a background state of the Universe where these processes could take place.
It has been demonstrated that incorporation of the  $S(T)$ 
within the SIV paradigm results in a compatible outcome with the standard BBNS
see the last two columns of Table  \ref{Table2}; 
furthermore, if one is to fit the observational data the result is $\lambda\approx1$
for the SIV parameter $\lambda$ (see last column of Table \ref{Table2} with $\lambda=\text{FRF}\approx1$). 
However, a pure SIV treatment (the middle three columns) results in $\Omega_b\approx1\%$ and less total matter,
either around $\Omega_\mathrm{m}\approx23\%$  when all the $\lambda$-scaling connections are utilized (see Table  \ref{Table2} column 6), 
or around $\Omega_\mathrm{m}\approx6\%$ without any $\lambda$-scaling factors 
(see column 5). 
The need to have $\lambda$ close to 1 is not an indicator of dark matter content but
indicates the goodness of the standard PRIMAT results that allows only for $\lambda$ close to 1 
as an augmentation, as such this leads to a light but important improvement in D/H 
as seen when comparing columns three with eight and nine.

The SIV paradigm suggests specific modifications to the reaction rates, 
as well as the functional temperature dependences of these rates,
that need to be implemented to have consistence between the 
EGR frame and the WIG (SIV) frame.
In particular, the non-in-scalar factor $T^\beta$ in the reverse reactions rates 
may be affected the most due to the SIV effects. 
As shown in \cite{VG&AM'23}, the specific dependences studied,   
within the assumptions made within the SIV model, resulted in three times less baryon matter,
usually around $\Omega_b\approx1.6\%$  and less total matter - around $\Omega_\mathrm{m}\approx6\%$.
The lower baryon matter content leads to also a lower photon to baryon ratio $\eta_{10}\approx2$
within the SIV, which is three tines less that the standard value of $\eta_{10}=6.09$.
As shown in \cite{VG&AM'23}, the overall results indicated insensitivity to the specific 
$\lambda$-scaling dependence of the \text{m\v{T}}-factor in the reverse reaction expressions within $T^\beta$ terms. 
Thus, one may have to explore further the SIV-guided $\lambda$-scaling relations as done
for the last column in  Table  \ref{Table2}, however, this would require the 
utilization of the numerical methods used by PRIMAT and as such will take us away from the 
SIV-analytic expressions explored that provided a simple model for understanding the BBNS within the SIV paradigm. 
Furthermore, it will take us further away from the accepted local statistical equilibrium and
may require the application of the reparametrization paradigm that seems to  result in SIV like 
equations but does not impose a specific form for $\lambda$ \cite{sym13030379}.
Thus, at this point the SIV theory is still a viable alternative model for cosmology.

\subsection{SIV and the Inflation of the Early Universe \cite{SIV-Inflation'21}}

Another important result within the SIV paradigm is the presence of inflationary stage at the very early Universe
$t\approx t_\text{in}\ll t_0=1$ with a natural exit from inflation in a later time $t_\mathrm{exit}$ with 
value related to the parameters of the inflationary potential  \cite{SIV-Inflation'21}.
The main steps towards these results are outlined below.

If we go back to the general scale-invariant cosmology Equation \eqref{E1p},
we can identify a vacuum energy density expression that relates 
the Einstein cosmological constant with the energy density
as expressed in terms of $\kappa=-\dot{\lambda}/\lambda$ by using the SIV result 
\eqref{SIV-gauge}. The corresponding vacuum energy density $\rho$, with $C=3/(4\pi G)$, is then:
\begin{equation}
\rho=\frac{\Lambda}{8\pi G}=\lambda^{2}\rho'=\lambda^{2}\frac{\Lambda_{E}}{8\pi G}
=\frac{3}{8\pi G}\frac{\dot{\lambda}^{2}}{\lambda^{2}}
=\frac{C}{2}\dot{\psi}^{2}\,.
\label{rho}
\end{equation}

\noindent
This provides a natural connection to inflation within the SIV 
via $\dot{\psi}=-\dot{\lambda}/\lambda$ or $\psi\propto\ln(t)$. 
The equations for the energy density, pressure, and Weinberg's condition for inflation 
within the standard model for inflation by \cite{Guth81, Linde95, Linde05, Weinberg08} are:
\begin{eqnarray}
\left.\begin{array}{c}\rho\\ p \end{array}\right\} 
=\frac{1}{2}\dot{\varphi}^{2}\pm V(\varphi),\label{rp}\;
\mid\dot{H}_{\mathrm{infl}}\mid\,\ll H_{\mathrm{infl}}^{2}\,.
\label{cond}
\end{eqnarray}

\noindent
If we make the identification between the standard model for inflation above with the 
fields present within the SIV (using $C=3/(4\pi G)$):
\begin{eqnarray}
\dot{\psi}=-\dot{\lambda}/\lambda, & \varphi\leftrightarrow\sqrt{C}\,\psi, \quad
V\leftrightarrow CU(\psi), & U(\psi)\,=\,g\,e^{\mu\,\psi}\,.
\label{SIV-identification}
\end{eqnarray}

\noindent
Here, $U(\psi)$ is the inflation potential with strength $g$ and field ``coupling'' $\mu$. 
One can evaluate the Weinberg's condition for inflation \eqref{cond} 
within the SIV framework \cite{SIV-Inflation'21}, and the result is:

\begin{equation}
\frac{\mid\dot{H}_{\mathrm{infl}}\mid}{H_{\mathrm{infl}}^{2}}\,
=\,\frac{3\,(\mu+1)}{g\,(\mu+2)}\,t^{-\mu-2}\ll1\,\text{ for }\ \mu<-2,\text{ and }\ t\ll t_{0}=1.
\label{crit}
\end{equation}

When the Weinberg's condition for inflation \eqref{cond} is not satisfied anymore, 
one can see that there is a graceful exit from inflation at the \mbox{later time:}
\begin{equation}
t_\mathrm{exit}\approx\sqrt[n]{\frac{n\ g}{3(n+1)}}\qquad \mathrm{with} \qquad n=-\mu-2>0.
\label{t_exit}
\end{equation}

The derivation of the equation \eqref{crit} starts with the use of 
the scale invariant energy conservation equation within SIV \cite{Maeder17a,SIV-Inflation'21}:
\begin{equation}
\frac{d(\varrho a^{3})}{da}+3\,pa^{2}+(\varrho+3\,p)\frac{a^{3}}{\lambda}\frac{d\lambda}{da}=0\,,
\label{conserv}
\end{equation}
which has the following equivalent form:
\begin{equation}
\dot{\varrho}+3\,\frac{\dot{a}}{a}\,(\varrho+p)+\frac{\dot{\lambda}}{\lambda}\,(\varrho+3p)\,=\,0\,.
\label{conserv2}
\end{equation}

\noindent
By substituting the expressions for $\rho$ and $p$ from \eqref{rp} 
along with the SIV identification \eqref{SIV-identification} 
within the SIV expression (\ref{conserv2}), one obtains modified form of the Klein--Gordon equation,
which could be non-linear when using non-linear potential $U(\psi)$ as in \eqref{SIV-identification}:
\begin{equation}
\ddot{\psi}+U'\,+3H_{\mathrm{infl}}\,\dot{\psi}-2\,(\dot{\psi}^{2}-U)\,=\,0\,.
\label{cons1}
\end{equation}

\noindent
The above Equation \eqref{cons1} can be used to evaluate the time derivative of the Hubble parameter.
The process is utilizing \eqref{SIV-gauge}; that is, 
$\lambda=t_{0}/t,\;\dot{\psi}=-\dot{\lambda}/\lambda=1/t\;\Rightarrow \ddot{\psi}=-\dot{\psi}^{2}$
along with $\psi=\ln(t)+const$ and $U(\psi)\,=\,g\,e^{\mu\,\psi}=g t^{\mu}$ when the normalization of the 
field $\psi$ is chosen so that $\psi(t_0)=\ln(t_0)=0$ for $t_0=1$ at the current epoch. The final result is: 
\begin{eqnarray}
H_{\mathrm{infl}}=\dot{\psi}-\frac{2\,U}{3\,\dot{\psi}}-\frac{U'}{3\,\dot{\psi}} &=&\,\frac{1}{t}-\frac{(2+\mu)\,g}{3}\,t^{\mu+1}\,,
\label{hh}\\
\dot{H}_{\mathrm{infl}}=-\dot{\psi}^{2}-\frac{2U}{3}-U'-\frac{U''}{3} &=&-\frac{1}{t^{2}}-\frac{(\mu+2)(\mu+1)\,g}{3}t^{\mu}\,.
\label{hd}
\end{eqnarray}

\noindent
For $\mu<-2$ the $t^{\mu}$ terms above are dominant; 
thus, the critical ratio \eqref{cond} for the occurrence of inflation near $t\approx t_\text{in}$ is then:
\begin{equation*}
\frac{\mid\dot{H}_{\mathrm{infl}}\mid}{H_{\mathrm{infl}}^{2}}\,=\,\frac{3\,(\mu+1)}{g\,(\mu+2)}\,t^{-\mu-2}\,.
\end{equation*}

Based on \eqref{rho}, both  $\varrho$ and $ \Lambda$ (in the scale invariant space) behave
like $1/t^2$ according to expression \eqref{lambda(t)} based on the field equation of the vacuum. 
This implies  that the energy density of the vacuum, and the cosmological constant $\Lambda$, 
in the scale invariant space become very large near the origin. 
For example, at the Planck time $t_{\mathrm{Pl}} = 5.39 \cdot 10^{-44}$ \, s, 
dominated by quantum effects, the cosmological constant would be a factor  
$\left (\frac{4.355 \cdot 10^{17}}{ 5.39 \cdot 10^{-44}}\right)^2 = \, 
6.4 \cdot 10^{121}$ larger than the  value at the present  cosmic age $\tau_0 = 13.7 \; \mathrm{Gyr} = 4.323 \cdot 10^{17}$ s.
Thus, as such this may solve the so-called cosmological problem by viewing the 
Planck-seed universes and the derivable universes as different stages of the same Universe
rather than a disconnected universe \cite{GueorM20}.
In other words, the smallness of the Einstein cosmological constant $\Lambda_E$  
is naturally related to the current age of the Universe, assuming that now $\lambda=1$ by choice of units,
because the solution  \eqref{lambda(t)} for \eqref{SIV-gauge} implies 
$\Lambda_E =3/\tau_0^2\approx 1.6\times10^{-35}\mathrm{s}^{-2}$.

\section{Conclusions and Outlook}

The SIV hypothesis is a relatively new theory, and it is still under development. 
However, the results of the tests that have been conducted so far are promising. 
If the SIV hypothesis is correct, it could provide a new and important understanding of the universe.

Based on the previous sections on various comparisons, 
one can conclude that the \textit{SIV cosmology is a viable alternative to }$\Lambda$CDM.
In particular, \textit{the cosmological constant disappears} within the SIV gauge (\ref{E1}). 
As emphasized in the discussion of Figure~\ref{rates} from \cite{Maeder17a}, 
there are diminishing differences in the values of the scale factor $a(t)$ within $\Lambda$CDM and SIV at higher densities. 
The SIV also shows consistency for $H_0$ and the age of the Universe, 
and the m-z diagram is well satisfied (see \cite{MaedGueor20a}  for details).

Furthermore,  \textit{the SIV provides the correct RAR for dwarf spheroidals} (Figure~\ref{gobs}), 
while this is more difficult for MOND, 
and even more, dark matter cannot yet account for the phenomenon \cite{MaedGueor20b}.
However, the observations still have some degree of uncertainty. 
What is clear is that,  as in other cases, within the SIV dark matter is not needed.

Thus, it seems that \textit{within the SIV, dark matter is not needed to seed the growth of structure} in the Universe, 
as there is a fast enough growth of the density fluctuations as seen in (Figure~\ref{variousn}) and 
discussed in more detail by \cite{MaedGueor19}.

In our study on inflation within the SIV cosmology,  we identified a connection between $\lambda$ and its rate of change, 
we have identified a connection of the scale factor $\lambda$, and its rate of change, 
$\dot{\psi}=-\dot{\lambda}/\lambda$ \eqref{SIV-identification},
to the conventional inflation field $\varphi$, that is, $\psi \rightarrow \varphi$.
As seen from \eqref{crit}, \textit{inflation of the very-very early Universe,  $\tau\approx0\; (t\approx\,t_{\text{in}}\ll1)$,
is natural, and SIV predicts a graceful exit from inflation} (see \eqref{t_exit})!

Our latest study on the primordial nucleosynthesis within the SIV \cite{VG&AM'23} 
has shown that smaller than usual baryon and non-zero dark matter content, 
by a factor of three to five times reduction, 
result in compatible to the standard light elements abundances (Table \ref{Table2}).

Some of the obvious future research directions are related to the primordial nucleosynthesis,
where preliminary results show a satisfactory comparison between SIV and observations \cite{Maeder19,VG&AM'23}. 
Further investigations of potential SIV-guided deviations from the local statistical equilibrium should be studied
since this may lead to mechanisms for understanding the  matter-anti-matter asymmetry. 
The recent success of the R-MOND in the description of the CMB \cite{2021PhRvL.127p1302S}, 
after the initial hope and concerns \cite{Skordis2006}, is very stimulating;
it suggests that a generally covariant theory that has the correct Newtonian limit 
is likely to describe the CMB; 
Since SIV is generally covariant and has the correct limits,
it demands testing the SIV cosmology as well 
against the MOND and $\Lambda$CDM successes in the description of the CMB, 
the Baryonic Acoustic Oscillations, etc.

Another area of research is to better understand the physical meaning of $\lambda$. 
As it was pointed out in Section 1.2, 
a general conformal factor seems to be linked to Jordan–Brans–Dicke scalar-tensor theory, 
resulting into a varying Newton's constant $G$ that has not been detected yet. 
Furthermore, a general conformal factor with spatial dependence opens the door to local field excitations that 
should manifest as some type of fundamental scalar particles. The Higgs boson is such a particle, 
but a connection to Jordan–Brans–Dicke scalar-tensor theory seems a far-fetched idea. 
On the other hand, the assumption of isotropy and homogeneity of space forces $\lambda$ 
to depend only on time, which is not in any sense similar to the familiar fundamental fields.
Thus, the specific time dependence for $\lambda(t)=t_0/t$ and the absence of  spatial dependence
are important hallmarks of the SIV paradigm. As such, they have to be maintained and appropriately derived from the
basic tenets of the SIV theory towards its future applications.

Furthermore,  some less obvious directions are: the solar system  exploration due to the 
high-accuracy data there, or exploring in more detail the connection to the re-parametrization invariance. 
In particular, it is known by \cite{sym13030379} that un-proper time parametrization  
leads to equations of motion \eqref{eq:geodesics+} corresponding with the weak-field SIV limit \eqref{Nvec}.

\vspace{6pt}

\paragraph*{Acknowledgment.} 
A.M. expresses his gratitude to his wife for her patience and support. 
V.G. is extremely grateful to his wife and daughters for their understanding and 
family support  during the various stages of the research presented. 
This research did not receive any specific grant from funding agencies in the public, 
commercial, or not-for-profit sectors.

\end{document}